\documentclass[pra,aps,showpacs,preprint]{revtex4}
\usepackage[latin1]{inputenc}
\usepackage{eufrak}
\usepackage{amsmath}
\usepackage{color}
\usepackage[dvips]{graphicx}
\usepackage{subfigure}

\def\/{\over}
\def\<{\left\langle}
\def\>{\right\rangle}
\def\({\left(}
\def\){\right)}
\def\[{\left[}
\def\]{\right]}
\def\d{{\rm d}}
\def\D{{\rm D}}
\def\im{{\rm i}}
\def\e{{\rm e}}

\def\bfgamma{\boldsymbol{\gamma}}

\def\bfrho{\boldsymbol{\rho}}

\def\O{\hbox{O}}

\def\Ai{\hbox{Ai}}
\def\det{\hbox{det}\,}

\def\sgn{\hbox{sgn}\,}
\def\arccot{\hbox{arccot}}

\begin{document}
\title{Semiclassical propagation of Wigner functions}
\author{T.~Dittrich$^{1,2}$, E.~A.~G\'omez$^{1,2}$,
L.~A.~Pach\'on$^{1,2}$}
\affiliation{$^{1}$Departamento de F\'{\i}sica, Universidad Nacional
de Colombia, Bogot\'a D.C., Colombia.\\
$^{2}$CeiBA -- Complejidad, Bogot\'a D.C., Colombia.}
\date{\today}

\begin{abstract}
We present a comprehensive study of semiclassical phase-space
propagation in the Wigner representation, emphasizing numerical
applications, in particular as an initial-value representation. Two
semiclassical approximation schemes are discussed: The propagator of
the Wigner function based on van Vleck's approximation replaces the
Liouville propagator by a quantum spot with an oscillatory pattern
reflecting the interference between pairs of classical
trajectories. Employing phase-space path integration instead, caustics
in the quantum spot are resolved in terms of Airy functions. We apply
both to two benchmark models of nonlinear molecular potentials, the
Morse oscillator and the quartic double well, to test them in
standard tasks such as computing autocorrelation functions
and propagating coherent states. The performance of
semiclassical Wigner propagation is very good even in the presence of
marked quantum effects, e.g., in coherent tunneling and in propagating
Schr\"odinger cat states, and of classical chaos in four-dimensional
phase space. We suggest options for an effective numerical
implementation of our method and for integrating it in
Monte-Carlo--Metropolis algorithms suitable for high-dimensional
systems.
\end{abstract}
\maketitle
\section{Introduction}
Molecular dynamics, by the spatial and temporal scales it involves,
straddles the borderline between quantum and classical
behavior. Quantum effects---above all the very concept of chemical
reaction---are obviously crucial. Notwithstanding, a good deal of a
molecule's external and even internal motion can be understood 
on purely classical
grounds: precisely the situation for which semiclassical methods have
been conceived and are optimally suited. Moreover, recent developments
in chemical physics, experimental as well as theoretical, are
advancing vigorously in the direction of complex time evolution and
high excitations. This faces adiabatic techniques like even
time-dependent density-functional theory (TDDFT) \cite{RG84} and
time-dependent Hartree \cite{GBR82,MMC90} with formidable challenges
while it favors semiclassical approaches which do not suffer from
limitations in this respect.

Therefore, the modest renaissance semiclassics in the time domain (for
a comprehensive review see \cite{SG96}) is presently enjoying
comes as no surprise. It includes 
approaches in configuration space as well as less
established phase-space methods. The well-known shortcomings
plagueging semiclassics in coordinate representation (based on WKB
approximations \cite{BM72} and the van-Vleck--Gutzwiller propagator
\cite{Vle28,Gut67,Lit92}), such as divergences at
caustics and the root-search problem, have largely been overcome by
sophisticated refinements of the original method
\cite{Mil74}. Remarkably, most of them already switch internally to
mixed representations (combining position with momentum) or full
phase space in order to smooth out divergences and to reduce the set
of contributing classical solutions to the neighbourhood of a single
trajectory, permitting the construction of initial-value
representations (IVRs): uniform approximations
\cite{LS77,LM&78,Kla86}, semiclassical IVRs
\cite{Mil74,Mil98,TWM01,LM06} (for reviews see
\cite{Mil01,TW04,Kay05}), Gaussian-wavepacket propagation \cite{Hel75}
with its numerous ramifications, notably the Herman-Kluk propagator
\cite{HK84,Gro99,BA&01} and Heller's cellular dynamics \cite{Hel91b}.
A similar but independent approach, the dephasing representation
\cite{Van04,Van06,LMV09} employs the shadowing lemma for chaotic
dynamics to represent the classical flow through a Planck cell by a
single trajectory in a semiclassical approximation for correlation
functions.

These methods have gained widespread acceptance in practical
applications but tend to be technically cumbersome. Their
implementation in \textit{ab initio} molecular-dynamics simulation
with ``on the fly'' update of the electronic evolution
\cite{MTM99,DM02a} has been proposed \cite{CA&09}. It has to
cope, though, with the high computational costs of calculating
Hessians for the potential-energy surface of the electron sector---so
much so that recourse is sought even at a complete suppression of
determinantal prefactors \cite{TP09}.

Semiclassical approximations working directly in phase space promise a
fresh, structurally more transparent approach. They are inherently
free of the above problems and thus offer a number of
tempting advantages: (i) phase-space propagation provides a natural
exact IVR by construction, without Gaussian smoothing, (ii) it can be
formulated exclusively in terms of canonically invariant quantities
(no projection required) that (iii) allow for simple geometrical
interpretations, (iv) as far as determinantal
prefactors arise, they are also canonically invariant, and (v) being
based on the density operator, not on wavefunctions, the extension to
non-unitary time evolution is immediate, opening access to decoherence
and dissipation \cite{ORB09}.

Yet there is a price to be paid. The two representations
mainly considered as candidates for semiclassical phase-space
propagation both come with their specific virtues and vices: The
Husimi function \cite{Hus40} is based on coherent states \cite{KS85},
hence closely related to semiclassical IVRs and Gaussian-wavepacket
propagation. Coherent states possess a natural interpretation in terms
of classical probabilities but are overcomplete, contain arbitrary
parameters, and, in order to work optimally, require to complexify
phase space \cite{Kla79,BA&01}. The Wigner function
\cite{Wig32,HO&84,Sch01}, by contrast, provides a parameter-free
one-to-one representation of the full density operator but encodes
information on quantum coherences as small-scale (``sub-Planckian'')
oscillatory fringes \cite{Ber77,Zur01}. They prevent a probability
interpretation and pose serious problems for their propagation,
pointed out more than 30 years ago by Heller \cite{Hel76}.

Owing to these seemingly prohibitive difficulties, the semiclassical
propagation of Wigner functions still awaits being explored in its
full potentiality. For a long time, it has been almost synonymous to a
mere classical propagation of phase-space distributions, discarding
quantum effects in the time evolution altogether \cite{Mcl83,GH95}. As
a first attempt to improve on this so-called classical Wigner
model, it suggests itself to include higher-order terms in $\hbar$
in the quantum analogue of the Poisson bracket, the Moyal bracket
\cite{Ber77}, upon integrating the evolution equation for the Wigner
function. Such additive quantum corrections give rise to
accordingly modified ``quantum trajectories''
\cite{LS82,Lee92,Raz96,DM01,THW03}. However, they tend to become
unstable even for short propagation time \cite{MP09} and suffer from
other practical and fundamental problems \cite{MS96}: As we shall
point out below, the very concept of propagating along single
deterministic trajectories is inadequate even in the semiclassical
regime.

A more convincing alternative has emerged in the context of quantum
chaos where semiclassical Wigner functions \cite{Ber77,Hel77,Ozo98}
are appreciated as valuable tools, e.g., in the study of scars in
wavefunctions and of spectral features related to them
\cite{Ber89,AF98,TAO01}. By applying semiclassical
approximations directly to the finite-time propagator, expanding
the phase instead of the underlying evolution equation, an
appropriate phase-space propagation method for semiclassical
Wigner functions has been achieved \cite{RO02}. It features, in a
geometrically appealing manner, \textit{pairs} of (real) classical
trajectories and the symplectic area enclosed between them as their
classical underpinning---not surprisingly in view of the
well-known pairing of paths and diagrams, one forward, one backward
in time, as it occurs in the propagation of quantities bilinear in the
state vector, particularly the density operator.

While this technique is tailor-made to time-evolve the specific class
of semiclassical Wigner functions \cite{Ber77,Hel77,Ozo98,TAO01},
semiclassical approximations to the propagator proper, independent of
the nature of initial and final states, have been constructed
\cite{DVS06} on basis of phase-space path integrals \cite{Mar91} and
of the Weyl-transformed van Vleck propagator \cite{Ber89}. This
approach opens the door towards a far broader range of applications.
It has already proven fruitful in an analysis of the spectral form factor
of classically chaotic systems in terms of phase-space manifolds
\cite{DP09}.

Here, by difference,
we pretend to demonstrate its viability as a practical instrument for
the propagation of molecular systems, including in particular 
numerical simulations. We therefore focus on its performance in
benchmark models established in chemical physics like the
Morse oscillator and the quartic double well and study standard tasks
such as propagating Gaussians and computing autocorrelation
functions. We are confident that the remarkable accuracy and
robustness of the scheme and its unexpected numerical stability
and efficiency suggest it as a new competitive option in semiclassical
propagation.

In Sec.~\ref{derivation}, we introduce the basic building blocks of
our method, specify our two main approaches, via the van-Vleck
propagator and via phase-space path integration, and
discuss general properties of the semiclassical Wigner propagator,
such as its geometrical structure and its asymptotics towards the
classical limit and towards weak anharmonicity. This section partially
corresponds to a more detailed account of material already published
in Ref.~\cite{DVS06}. Readers mainly interested in numerical results
might proceed directly to Sec.~\ref{numerics}, dedicated to a
broad survey of the performance of semiclassical Wigner propagation
for prototypical models of nonlinear molecular potentials, in the
computation of dynamical functions like autocorrelations as well as of
detailed structures in phase space. We also test our method in
particularly challenging situations, involving strong quantum
effects, such as the reproduction of coherent tunneling and the
propagation of Schr\"odinger-cat states, or a classically chaotic
dynamics. Strategies to optimize the implementation of this method in
numerical, including Monte-Carlo-like, algorithms are suggested in
Sec.~\ref{algorithms}. We conclude in Sec.~\ref{conclusion}
with an outlook to immediate extensions of our work.

\section{Constructing the semiclassical Wigner propagator}
\label{derivation}

\subsection{Definitions and basic relations}
\label{definitions}

Alternative to its definition as a Weyl-ordered transform of the
density operator \cite{HO&84}, the Wigner function can be introduced
as the expectation value of phase-space reflection operators
\cite{Ozo98} or as expansion coefficient function in a basis of
phase-space displacement operators \cite{GT88}, besides other
options. For our purposes, the standard definition suffices,
\begin{equation}\label{wigfunc}
W({\bf r}) = \frac{1}{2\pi\hbar}\int\d^f q'\,
\e^{-\im{\bf p\cdot q'}/\hbar} \<{\bf q} + \frac{{\bf q}'}{2}\right|
\hat\rho\left|{\bf q} - \frac{{\bf q}'}{2}\>,
\end{equation}
where $\hat\rho$ denotes the density operator and ${\bf r} = ({\bf
p},{\bf q})$ is a vector in $2f$-dimensional phase space (we adopt
this ordering throughout the paper but assign $q$
to the horizontal and $p$ to the vertical axis in phase-space plots).

We shall restrict ourselves to unitary time
evolution within the dynamical group $\hat U(t) =
\exp(-\im\hat H t/\hbar)$ generated by a time-independent
Hamiltonian $\hat H$. The extension to time-dependent Hamiltonians is
immediate but will be suppressed to avoid technicalities. The
von-Neumann equation for the density operator, $\d\hat\rho/\d t =
(-\im/\hbar)[\hat H,\hat\rho]$, translates into an equation of motion
for the Wigner function, $(\partial/\partial t) W({\bf r},t) =
\{H_{\rm W}({\bf r}),W({\bf r},t)\}_{\rm Moyal}$, involving the Weyl
symbol $H_{\rm W}({\bf r})$ of the Hamiltonian $\hat H$. For $\hbar\to
0$, the Moyal bracket $\{\cdotp,\cdotp\}_{\rm Moyal}$
\cite{HO&84,Sch01} converges to the Poisson bracket.

The time evolution of the Wigner function over a finite time can be
expressed as an integral kernel, the Wigner propagator $G_{\rm W}({\bf
r}'',{\bf r}',t)$,
\begin{equation}\label{wigprop}
W({\bf r}'',t) = \int \d^{2f}r'\,G_{\rm W}({\bf r}'',{\bf r}',t)
W({\bf r}',0).
\end{equation}
It forms a one-parameter group, implying in particular the
initial condition $G_{\rm W}({\bf r}'',{\bf r}',0) = \delta({\bf r'' -
r'})$ and the composition law
\begin{equation}\label{wigconcat}
G_{\rm W}({\bf r}'',{\bf r}',t) =
\int \d^{2f}r\, G_{\rm W}({\bf r}'',{\bf r},t-t')
G_{\rm W}({\bf r},{\bf r}',t').
\end{equation}

An important quantity that must not be confused with the Wigner
propagator is the Weyl transform of the evolution operator $\hat
U(t)$, referred to as the Weyl propagator for short \cite{Ozo98},
\begin{equation}\label{weylprop}
U_{\rm W}({\bf r},t) = \int\d^f q'
\e^{\frac{-\im}{\hbar}{\bf p}\cdot{\bf q}'}
\<{\bf q}+{{\bf q}'\/2}\right|\hat U(t)
\left|{\bf q}-{{\bf q}'\/2}\>.
\end{equation}
It cannot be used to propagate Wigner functions as it stands, but is
related to the Wigner propagator by a self-convolution
\cite{Mar91,RO02,DVS06},
\begin{equation}\label{wigweyl}
G_{\rm W}({\bf r}'',{\bf r}',t) =
\frac{1}{(2\pi\hbar)^{2f}} \int \d^{2f}R\,
\e^{\frac{-\im}{\hbar}({\bf r}''-{\bf r}')\wedge{\bf R}}
U_{\rm W}^*(\tilde{\bf r}_-,t)U_{\rm W}(\tilde{\bf r}_+,t),
\end{equation}
with $\tilde{\bf r}_{\pm} \equiv ({\bf r}'+{\bf r}'' \pm {\bf
R})/2$. It serves as a starting point for the construction of
semiclassical approximations, based on the van-Vleck propagator
(Sec.~\ref{vanvleck}) as well as on phase-space path
integration (Sec.~\ref{marinov}).

\subsection{van Vleck approximation for the Wigner propagator}
\label{vanvleck}
\begin{figure}[floatfix]
\includegraphics[scale=0.3]{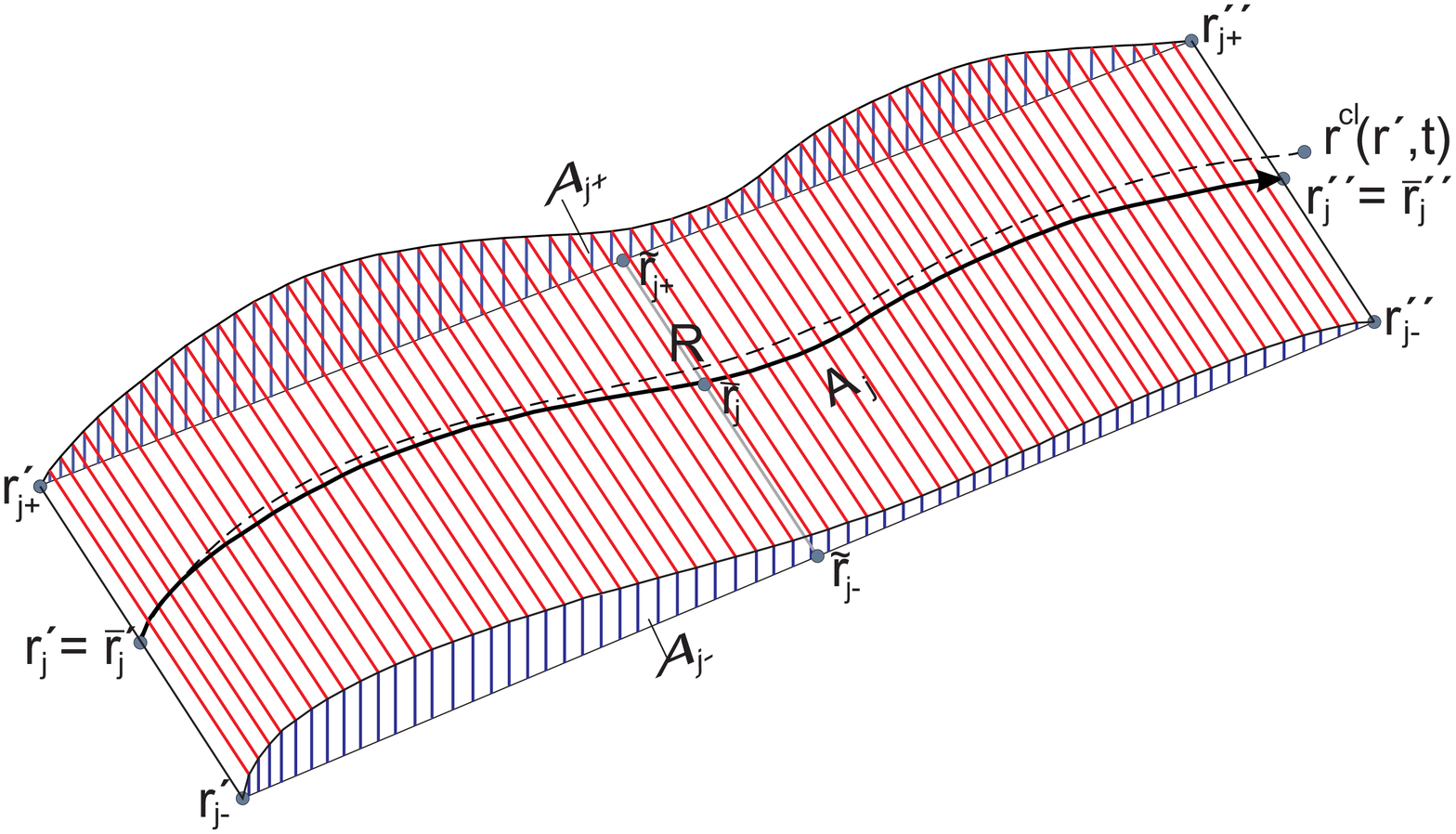}
\caption{\label{ainama} 
Symplectic areas entering the van-Vleck-based semiclassical Wigner
propagator, Eqs.~(\protect\ref{wigvleckprop},
\protect\ref{wigvleckaction}). The vertically hashed areas
correspond to the phases $A_{j_\pm}$ of the Weyl propagators
(\protect\ref{weylvleck}) according to the chord rule. The symplectic
area (slanted hatching) enclosed between the two classical
trajectories ${\bf r}_{j\pm}(t)$ and the two transverse vectors ${\bf
r}'_{j+} - {\bf r}'_{j-}$ and ${\bf r}''_{j+} - {\bf r}''_{j-}$
determines the phase (\protect\ref{wigvleckaction}) of the
propagator. The classical trajectory ${\bf r}^{\rm cl}({\bf r}',t)$
(dashed) is to be distinguished from the propagation path $\bar{\bf
r}_j({\bf r}',t)$ (bold) connecting the initial argument ${\bf r}'$ of
the propagator to the final one, ${\bf r}''$.
}
\end{figure}

A straightforward route towards a semiclassical Wigner propagator is
replacing the Weyl propagator in Eq.~(\ref{wigweyl}) by the Weyl
transform of the van Vleck propagator
\cite{Ber89,Mar91,Ozo98}. Transformed from the energy to the time
domain, it reads \cite{Mar91,Ozo98},
\begin{equation}\label{weylvleck}
U_{\rm W}({\bf r},t) = 2^f \sum_j \frac{\exp\({\im\/\hbar}
S_j({\bf r},t) - \im\mu_j\frac{\pi}{2}\)}
{\sqrt{|\det(\mathsf{M}_j({\bf r},t)+\mathsf{I})|}}.
\end{equation}
The sum runs over all classical trajectories $j$ connecting
phase-space points ${\bf r}'_j$ to ${\bf r}''_j$ in time $t$ such that
${\bf r} = \tilde{\bf r}_j \equiv ({\bf r}'_j+{\bf r}''_j)/2$
(midpoint rule). $\mathsf{M}_j$ and $\mu_j$ are its stability
matrix and Maslov index, respectively. The action $S_j({\bf r}_j,t) =
A_j({\bf r}_j,t) - H_j({\bf r},t)\,t$, with $H_j({\bf r},t) \equiv
H_{\rm W}({\bf r}_j,t)$, the Weyl Hamiltonian evaluated on the
trajectory $j$ (to be distinguished from $H_{\rm W}({\bf r},t)$) and
$A_j$, the symplectic area enclosed between the trajectory and the
straight line (chord) connecting ${\bf r}'_j$ to ${\bf r}''_j$
\cite{Ber89} (chord rule, vertically hashed areas $A_{j_\pm}$ in
Fig.~\ref{ainama}).

Substituting Eq.~(\ref{weylvleck}) in (\ref{wigweyl}), one arrives at
\begin{align}\label{wigvleck}
& \quad G_{\rm W}({\bf r}'',{\bf r}',t) = 
\frac{1}{(\pi\hbar)^{2f}}\sum_{j_-,j_+} \int \d^{2f}R\,
\e^{\frac{-\im}{\hbar}({\bf r}''-{\bf r}')\wedge{\bf R}}
\nonumber\\
& \!\!\!\!\!\frac{\exp\[\frac{\im}{\hbar}
\(S_{j_+}(\tilde{\bf r}_{j_+},t) - S_{j_-}(\tilde{\bf r}_{j_-},t)\) -
\im(\mu_{j_+}-\mu_{j_-})\frac{\pi}{2}\]}
{|\det[\mathsf{M}_{j_-}({\bf r},t)+\mathsf{I}]\,
\det[\mathsf{M}_{j_+}({\bf r},t)+\mathsf{I}]|^{1/2}},
\end{align}
where indices $j_\pm$ refer to classical trajectories contributing to
the Weyl propagators $U_{\rm W}(\tilde{\bf r}_\pm,t)$ in
Eq.~(\ref{wigweyl}). The principal challenge is now evaluating the
$R$-integration. As it stands, Eq.~(\ref{wigvleck}) couples the two
classical trajectories ${\bf r}_{j_-}$, ${\bf r}_{j_+}$,
to one another only indirectly via ${\bf R}$, the separation of
their respective midpoints. This changes as soon as an integration by
stationary-phase approximation is attempted, consistent with the
use of the van Vleck propagator. Stationary points are specified
implicitly by the condition ${\bf r}''-{\bf r}' = ({\bf r}''_{j-}-{\bf
r}'_{j-}+{\bf r}''_{j+}-{\bf r}'_{j+})/2$. Combined with the midpoint
rule ${\bf r}'+{\bf r}''\pm{\bf R} = {\bf r}'_{j\pm}+{\bf
r}''_{j\pm}$, this implies
\begin{equation}\label{midpoints}
{\bf r}' = \bar{\bf r}'_j \equiv
({\bf r}'_{j-} + {\bf r}'_{j+})/2,\quad
{\bf r}'' = \bar{\bf r}''_j \equiv
({\bf r}''_{j-} + {\bf r}''_{j+})/2.
\end{equation}
Equation (\ref{midpoints}) constitutes a simple geometrical rule for
semiclassical Wigner propagation \cite{Ozo98}: It is based on pairs of
classical trajectories $j_+$, $j_-$, that need not coincide with one
another nor with the trajectories passing through ${\bf r}'$ and ${\bf
r}''$ but must have ${\bf r}'$ midway between their respective
initial points ${\bf r}'_{j\pm}$ and likewise for ${\bf r}''$. This
condition is trivially fulfilled for identical pairs ${\bf
r}_{j_-}(t) = {\bf r}_{j_+}(t)$, as a type of diagonal
approximation. By including also non-diagonal terms, i.e.,
non-identical trajectory pairs, we add classical information contained
in third-order terms in the action and partially compensate for the
general limitations of the van Vleck approach on the level of
wavepackets. This becomes particularly evident in the case of pure
initial states $\hat\rho' = |\psi'\rangle\langle\psi'|$: Applying the
van Vleck propagator in position representation separately to ket and
bra, allowing for off-diagonal terms in the ensuing double sum over
classical orbits, and then transforming back to the final Wigner
function leads to the same result as the present derivation, albeit
through a more tedious and less transparent route.

To complete the Fresnel integral over $R$, we note that 
\begin{equation}\label{wigvleckdet}
\frac{\partial^2}{\partial{\bf R}^2}
\[S_{j_+}(\tilde{\bf r}_+,t) - S_{j_-}(\tilde{\bf r}_-,t)\] 
= \frac{\mathsf{J}}{2}\(\frac{\mathsf{M}_{j_-}-\mathsf{I}}
{\mathsf{M}_{j_-}+\mathsf{I}} -
\frac{\mathsf{M}_{j_+}-\mathsf{I}}
{\mathsf{M}_{j_+}+\mathsf{I}}\)
= \frac{\mathsf{J} (\mathsf{M}_{j_-}-\mathsf{M}_{j_+})}
{(\mathsf{M}_{j_-}+\mathsf{I})(\mathsf{M}_{j_+}+\mathsf{I})},
\end{equation}
where $\mathsf{J}$ denotes the $2f\times 2f$ symplectic unit matrix
\cite{Ber89}. Combined with the determinantal prefactors inherited
from the van Vleck propagator, this produces
\begin{equation}\label{wigvleckprop}
G_{\rm W}^{\rm vV}({\bf r}'',{\bf r}',t) = {4^f\/h^f}\sum_j
{2\cos\(\frac{1}{\hbar}S_j^{\rm vV}({\bf r}'',{\bf r}',t)-
f\frac{\pi}{2}\)\/
|\det(\mathsf{M}_{j_+}-\mathsf{M}_{j_-})|^{1/2}},
\end{equation}
our main result for the semiclassical Wigner propagator in van Vleck
approximation. The phase is determined by
\begin{align}\label{wigvleckaction}
& S_j^{\rm vV}({\bf r}'',{\bf r}',t) = 
(\tilde{\bf r}_{j_+}-\tilde{\bf r}_{j_-})
\wedge({\bf r}''-{\bf r}')+S_{j_+}-S_{j_-} \notag\\
& = \int_0^t \d s\,\big[{\dot{\bar{\bf r}}}_j(s)\wedge
{\bf R}_j(s)-H_{j_+}({\bf r}_{j_+})+H_{j_-}({\bf r}_{j_-})\big],
\end{align}
with $\bar{\bf r}_j(s) \equiv ({\bf r}_{j_-}(s) + {\bf
r}_{j_+}(s))/2$ and ${\bf R}_j(s) \equiv {\bf r}_{j_+}(s) - {\bf
r}_{j_-}(s)$. Besides the two Hamiltonian terms it includes the
symplectic area enclosed between the two trajectory sections and the
vectors ${\bf r}'_{j_+} - {\bf r}'_{j_-}$ and ${\bf r}''_{j_+} - {\bf
r}''_{j_-}$ (Fig.~\ref{ainama}).

\begin{figure}[floatfix]
 \includegraphics[scale=0.3]{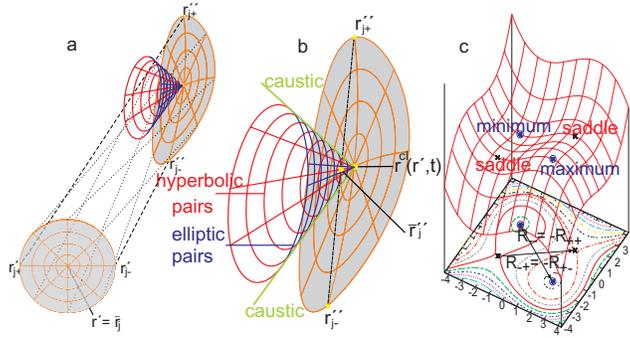}
\caption{\label{zanahoria} 
Classical skeleton of the semiclassical Wigner propagator according to
Eqs.~(\protect\ref{wigvleckprop},\protect\ref{wigvleckaction}). A set
of initial points ${\bf r}'_{j_\pm}$ (panel a, left target pattern)
surrounding the initial point ${\bf r}'$ of the propagation path is
propagated along classical trajectories (dotted lines), giving rise to
a corresponding set of final points ${\bf r}''_{j_\pm}$ (right
deformed target pattern). The manifold formed by their midpoints
(cone-like structure, panel b) constitutes the ``illuminated area'',
the support of the semiclassical propagator. Inside this area, two
trajectory pairs contribute to each point, one of them elliptic
(``upper'' shell of the cone, blue), the other hyperbolic (``lower''
shell, red). Its boundary gives rise to caustics in phase space where
the two trajectory pairs collapse into one. Panel c depicts the
generic structure of the underlying phase as a function of the
integration variable ${\bf R}$ in Eq.~(\protect\ref{wigvleck}), an odd
cubic polynomial of the phase-space coordinates with two extrema, a
minimum and a maximum (blue $\odot$ symbols) and two saddles (red
$\times$ symbols), corresponding to the elliptic and the hyperbolic
trajectory pairs, respectively.
}
\end{figure}

In the following we list a number of general features of
Eqs.~(\ref{wigvleckprop},\ref{wigvleckaction}):

\renewcommand{\theenumi}{\roman{enumi}}
\begin{enumerate}

\item Equation (\ref{wigvleckprop}) replaces the Liouville propagator,
\begin{equation}\label{liouprop}
G_{\rm W}^{\rm cl}({\bf r}'',{\bf r}',t) =
\delta\[{\bf r}'' - {\bf r}^{\rm cl}({\bf r}',t)\], 
\end{equation}
localized on the classical trajectory ${\bf r}^{\rm cl}({\bf
r}',t)$ initiated in ${\bf r}'$, by a ``quantum spot'', a smooth
distribution peaked at the support of the classical propagator but
spreading into the adjacent phase space and structured by an
oscillatory pattern that results from the interference of the
trajectories involved.

\item The propagator (\ref{wigvleckprop}) does involve determinantal
prefactors. However, they do not result from any projection onto a
subspace like ${\bf q}$ or ${\bf p}$ and are
invariant \cite{Ber89} under linear canonical (affine) transformations
\cite{Bal80}.

\item \label{holimit} It deviates from the Liouville propagator
only if the potential is anharmonic. For a purely harmonic potential,
the two operations, propagation in time and forming midpoints between
trajectories, commute, so that all midpoint paths $\bar{\bf r}_j(t) =
[{\bf r}_{j-}(t) + {\bf r}_{j+}(t)]/2$ coincide with each other
and with the classical trajectory ${\bf r}^{\rm cl}({\bf r}',t)$. This
singularity restores the classical delta function on ${\bf r}^{\rm
cl}({\bf r}',t)$, see Sec.~\ref{asymptotics}.

\item By contrast to position-space semiclassics \cite{Lit92}, the
number of trajectory pairs contributing to the summation in
Eq.~(\ref{wigvleckprop}) ranges between 0,
1, and 2 following a universal pattern (Fig.~\ref{zanahoria}):
Within an ``illuminated area'', a sector with its tip on the classical
trajectory, the sum contains two trajectory
pairs (four stationary points, Fig.~\ref{zanahoria}c). In the
``shadow region'' outside this sector, stationarity cannot be
fulfilled by real trajectories. Along the border
there is exactly one solution (two stationary points). Unexpected
in phase space, this caustic arises as the projection onto phase space
of the manifold of midpoints $\bar{\bf r}''_j = ({\bf r}''_{j-} + {\bf
r}''_{j+})/2$ (Fig.~\ref{zanahoria}b).

\item The set of trajectory pairs to enter the calculation of the
propagator is only restricted by the midpoint
rule (\ref{midpoints}). However, this does not constitute a
double-sided boundary condition since every pair fulfilling
Eq.~(\ref{midpoints}) for the initial points contributes a valid data
point to the propagator: \textit{no root-search}.
The freedom in the choice of trajectory pairs
can be exploited to optimize algorithms, see Sec.~\ref{vvprop}.

%
\item \label{ellhypshells} Of the two trajectory pairs contributing
to the illuminated area, exactly one is elliptic, associated to a
pair of opposite extrema of the action, one is hyperbolic
and associated to a pair of saddles (Fig.~\ref{zanahoria}c). They can
be distinguished by monitoring $\sgn\det(\mathsf{M}_{j_+} -
\mathsf{M}_{j_-})$, positive for the former and negative for the
latter. The two sets form separate sheets of the action
(\protect\ref{wigvleckaction}). Within each of them, amplitude and
phase of the propagator are slowly varying functions of ${\bf r}''$,
facilitating their numerical treatment, see Sec.~\ref{algorithms}.

%
\item The propagator's oscillatory pattern encodes information on
quantum coherences and allows us to propagate the ``sub-Planckian
oscillations'' that characterize the Wigner function, solving the
problem of the ``dangerous cross terms'' pointed out by Heller
\cite{Hel76}. See Sec.~\ref{schroedinger}.

\item The principle of propagation by trajectory pairs is consistent
with the properties of a dynamical group. It translates the
concatenation of propagators into a continuation of
trajectories, if the convolution (\ref{wigconcat}) is
evaluated by stationary phase as well.

\item Equations (\ref{wigvleckprop}, \ref{wigvleckaction}) fail if the
stationary points approach each other too closely. This is the case
near the central peak of the propagator on the classical trajectory
and along the caustics mentioned above. The same problem arises in the
limits $t\to 0$, $\hbar\to 0$, and of weak anharmonicity. It can be
overcome by means of a uniform approximation to the $R$-integration in
Eq.~(\ref{wigvleck}), see Sec.~\ref{perturbation}.

\item As a consequence of the underlying van Vleck approximation, the
propagator (\ref{wigvleckprop}) is not properly
normalized. Specifically, the slow decay of its oscillatory tail
renders it nonintegrable with respect to its initial or final
arguments, see Sec.~\ref{autocorrelation} for quantitative
details. The problem is solved by the same uniform approximation as
announced above.

\end{enumerate}

\subsection{Path-integral approach}
\label{marinov}

The quality of the van Vleck approximation (\ref{wigvleckprop}) to the
Wigner propagator is limited by its principal ingredient, the
van Vleck propagator itself. Just as in configuration-space
propagation, a comprehensive solution is provided on the basis of path
integrals \cite{Fey48,Sch81}. In phase space
an analogous theoretical framework is available since the
seminal work of Marinov \cite{Mar91}, largely following
Feynman: Time is discretized into $N$ equidistant sections $t_n
= t'+n\Delta t$, where $\Delta t = T/N$ and $T = t''-t'$ is the time
span to be propagated over. At each $t_n$, two short-time propagators,
composed of the corresponding Weyl propagators according to
Eq.~(\ref{wigweyl}), are concatenated by means of
Eq.~(\ref{wigconcat}). In the continuous limit $N \to \infty$, $\Delta
t \to 0$, the full propagator is obtained as a double path integral
comprising two distinct path variables ${\bf r}$ and ${\bf R}$,
inherited, respectively, from Eqs.~(\ref{wigconcat}) and
(\ref{wigweyl}), 
\begin{equation}\label{pathint}
G_{\rm W}({\bf r}'',t'';{\bf r}',t') = h^{-f}\int\D r\int\D R \,
\e^{\frac{-\im}{\hbar}S(\{{\bf r}\},\{{\bf R}\})}
\end{equation}
with an action integral
\begin{equation}\label{pathaction}
S(\{{\bf r}\},\{{\bf R}\}) = \int_{t'}^{t''}\d s
\biggl\{\dot{\bf r}(s)\wedge{\bf R}(s)
+H_{\rm W}\[{\bf r}(s)+\frac{1}{2}{\bf R}(s)\]-
H_{\rm W}\[{\bf r}(s)-\frac{1}{2}{\bf R}(s)\]\biggr\}.
\end{equation}
Only ${\bf r}(t)$ is subject to boundary conditions ${\bf r}(t') =
{\bf r}'$ and ${\bf r}(t'') = {\bf r}''$ while ${\bf R}(t)$ is free
and may be associated to quantum fluctuations.

For this subsection, we restrict ourselves to
a single degree of freedom, ${\bf r} = (p,q)$, and to standard
Hamiltonians $H(\hat p,\hat q,t) = T(\hat p) + V(\hat
q,t)$, $T(\hat p) = \hat p^2/2m$, but admit an explicit time
dependence and an arbitrary degree of nonlinearity of the
potential. In this case, the Weyl Hamiltonian, $H_{\rm
W}(p,q)$, is obtained by replacing the operators $\hat p$,
$\hat q$ in $H$ with corresponding real variables, $H_{\rm
W}(p,q) = T(p) + V(q,t)$.

Returning to discrete time, we are faced with evaluating the action
\begin{equation}\label{discaction}
S_N(\{{\bf r}\},\{{\bf R}\}) = \sum_{n=1}^{N}
\biggl\{\Delta{\bf r}_n\wedge{\bf R}_n
+ \[H_{\rm W}\(\bar{\bf r}_n+\frac{1}{2}{\bf R}_n\) -
H_{\rm W}\(\bar{\bf r}_n-\frac{1}{2}{\bf R}_n\)\]\Delta t\biggr\},
\end{equation}
denoting $\Delta{\bf r}_n \equiv {\bf r}_n - {\bf r}_{n-1}$ and
$\bar{\bf r}_n \equiv ({\bf r}_{n-1} + {\bf r}_n)/2$.
To obtain a semiclassical approximation, we expand the
action (\ref{discaction}) in ${\bf r}_n$ around a classical trajectory
${\bf r}_n^{\rm cl}$ defined implicitly in discrete time by
$\Delta p_n^{\rm cl} = -V'(\bar q_n^{\rm cl},t_n)\Delta t$,
$\Delta q_n^{\rm cl} = (\bar p_n^{\rm cl}/m)\Delta t$,
and around ${\bf R}_n \equiv {\rm 0}$, $n = 1,\ldots,N$. By expanding
around a single trajectory, we are sampling less classical information
than in the van Vleck approximation based on trajectory pairs. This
renders our path-integral approach inferior in some respects
detailed below. Note that the action (\ref{discaction}) is not only an
odd polynomial in ${\bf R}_n$, it is even lacking a linear term, hence
$\O(R^3)$, since according to Hamilton's equations of motion, the
first term $\sim ({\rm d}{\bf r}^{\rm cl}/{\rm d}t) \wedge {\bf
R}(t) = {\bf R}(t) \wedge \mathsf{J}\,\partial H_{\rm W}({\bf r}^{\rm
cl})/\partial {\bf r}^{\rm cl}$ cancels exactly the leading order of a
Taylor expansion of $H_{\rm W}({\bf r}^{\rm cl}+{\bf R}/2) - H_{\rm
W}({\bf r}^{\rm cl}-{\bf R}/2)$ with respect to ${\bf R}$. 

Expanding to third order in all variables (there is no $\O(R^4)$-term
in $S_N(\{{\bf r}\},\{{\bf R}\})$ anyway), this leaves us with
\begin{align}\label{truncaction}
&S_N(\{{\bf r}\},\{{\bf R}\}) = 
S_N(\{{\bf r}^{\rm cl}\},\{{\bf 0}\})
+ \sum_{n=1}^N
\big[\(\Delta q_n - T''(\bar p_n^{\rm cl})\bar p_n\Delta t\)P_n
\notag\\
&+ \(\Delta p_n + V''(\bar q_n^{\rm cl},t_n)\bar q_n\Delta t\)Q_n
+ \frac{1}{24}V'''(\bar q_n^{\rm cl},t_n)Q_n^3\big].
\end{align}
The $N$ integrations over the $R_n$ can now be performed, much like
the $p$-integrations in the derivation of the Feynman path integral
\cite{Fey48,Sch81}, resulting in one-step propagators
\begin{equation}\label{onestepprop}
G_{\rm W}({\bf r}_n,t_n;{\bf r}_{n-1},t_{n-1}) =
\delta\[\Delta q_n - T''_n p_{n-1}\Delta t\]
\nu_n^{-1/3}\Ai\[\nu_n^{-1/3}
\(\Delta p_n + V''_n q_{n-1}\Delta t\)\],
\end{equation}
where we use shorthands $T_n = T(\bar p_n^{\rm cl})$ and $V_n =
V(\bar q_n^{\rm cl},t_n)$ and define $\nu_n = \hbar^2 V'''_n \Delta t
/8$. The factor $\nu_n^{-1/3}$ is analogous to the determinantal
prefactors occurring in position-space path integration in that it
involves higher derivatives of the potential. Similarly as in the
van-Vleck-based Eq.~(\ref{wigvleckprop}), however, it constitutes a
mathematically more benign canonical invariant.

Taking into account that the one-step propagators
(\ref{onestepprop}) are already written in
a frame moving with the classical trajectory, we observe
three independent actions of the propagator on the evolving quantum
spot: (i), a rigid shift along the classical trajectory, (ii), a
rotation (shear) with the linearized phase-space flow around
the classical trajectory, if it is elliptic (hyperbolic), and (iii),
quantum Airy-function spreading in the negative $p$-direction
if $V'''(q)$ is positive and vice versa.
%
%
%
%
%
%
Their concatenation is
most conveniently performed in Fourier space, where, under certain
conditions fulfilled here \cite{fn1}, the convolution
(\ref{wigconcat}) reduces to a mere multiplication. With rescaled
coordinates of common dimension $\sqrt{\hbox{action}}$,
\begin{equation}\label{scalepq}
\bfrho = (\eta,\xi) = \big((V''/T'')^{-1/4}p,(V''/T'')^{1/4}q\big),
\end{equation}
the transformation of the Wigner propagator to Fourier phase space
$\bfgamma = (\alpha,\beta)$ is given as
\begin{equation}\label{fourierwigprop}
\tilde G_{\rm W}(\bfgamma'',t'';\bfgamma',t') = \frac{1}{4\pi^2}
\int\d\rho''^2\,\int\d\rho'^2\,
\e^{-\im(\bfgamma''\wedge\bfrho'' - \bfgamma'\wedge\bfrho')}
G_{\rm W}(\bfrho'',t'';\bfrho',t').
\end{equation}
Going to the continuum limit, we thus obtain
\begin{equation}\label{wigpropfourier}
\tilde G_{\rm W}^{\rm pi}(\bfgamma'',t'';\bfgamma',t') =
\delta\big[\bfgamma'' - \mathsf{M}(t'',t')\bfgamma'\big]
\exp\left\{\frac{-\im}{3}\int_{t'}^{t''}\d t\,
\sigma(t) \[\mathsf{M}(t,t')\bfgamma'\]_{\beta}\right\}.
\end{equation}
Here, $\mathsf{M}(t,t') = \partial{\bf r}^{\rm cl}(t)/\partial{\bf
r}^{\rm cl}(t')$ is the stability matrix of
${\bf r}^{\rm cl}(t)$, the subscript $\beta$ selects the second
component of $\bfgamma$, and
\begin{equation}\label{sigma}
\sigma(t) \equiv \frac{\hbar^2}{8}
\left|\frac{V''[q^{\rm cl}(t),t]}{T''[p^{\rm cl}(t)]}\right|^{-3/4}
V'''(q^{\rm cl}(t),t). 
\end{equation}
Inverting the Fourier transform (\ref{fourierwigprop}) we recover the
Wigner propagator. In the following subsection, we work
out this result in detail for the specific case $\sigma(t) \equiv
\mathrm{const}$.

\subsection{Weak anharmonicity, short-time and classical limits}
\label{specials}
\subsubsection{Path-integral approach for weak cubic nonlinearity}
\label{cubic}
Equation (\ref{wigpropfourier}) permits an analytical evaluation in
the case of constant second- and third-order derivatives of the
potential, which we will discuss in this subsection. It
can arise in different scenarios: To begin with, consider a static
binding potential with cubic nonlinearity, e.g., $V(q) =
(m\omega^2/2)q^2(1+q/3q_{\rm min})$, and a trajectory close to
the quadratic minimum at $q_{\rm min}$, so that the linearized
flow is elliptic. Then $\partial^3 V(q)/\partial q^3 \equiv V''' =
m\omega^2/q_{\rm min} = \mathrm{const}$ and we can also assume
$(V''/T'')^{1/2} \approx \omega = \mathrm{const}$. In accordance
with Eq.~(\ref{sigma}), denote $\sigma = (\hbar^2/8)
(m\omega)^{-3/2}V''' = (\hbar^2/8q_{\rm min})\sqrt{\omega/m}$. In
this case, the stability matrix corresponds to a rotation by $\tau =
\omega(t-t')$,
\begin{equation}\label{contmap}
\mathsf{M}(t,t') = \[\begin{array}{cc}\cos\,\tau&-\sin\,\tau\\
\sin\,\tau&\cos\,\tau\end{array}\],
\end{equation}
and the phase of the Fourier transformed propagator
(\ref{wigpropfourier}) reduces to elementary integrals,
\begin{equation}\label{wigpropcubicfourier}
\tilde G_{\rm W}^{\rm pi}(\bfgamma'',t'';\bfgamma',t') =
\delta\big[\bfgamma'' - \mathsf{M}(t'',t')\bfgamma'\big] \notag\\
\exp\[\frac{-\im\sigma}{3\omega}
\int_{\tau'}^{\tau''}\d\tau\, 
(\alpha'\sin\,\tau + \beta'\cos\,\tau)^3\].
\end{equation}
In a reference frame $\bar\bfrho = (\bar\eta,\bar\xi)$ rotating by
$\tau/2$ with respect to $\bfrho$,
$\bar\eta = \eta\cos(\tau/2) - \xi\sin(\tau/2)$,
$\bar\xi = \eta\sin(\tau/2) + \xi\cos(\tau/2)$,
the propagator after inverting the Fourier transform takes a
particularly transparent form 
\begin{equation}\label{wigpropcubicell}
G_{\rm W}^{\rm pi}(\bar\bfrho'',t'';\bar\bfrho',t') =
\pi \sqrt{f(s)} \kappa(s)^{-2/3}
\Ai[\kappa(s)^{-1/3}\bar\rho_-]
\Ai[\kappa(s)^{-1/3}\bar\rho_+],
\end{equation}
where we abbreviate $\kappa(s) = \sigma s^3 f(s)/3\omega$,
$f(s) = 3s^{-2}-1$, $s = \sin(\tau/2)$, and
$\bar\rho_\pm = \(\bar\eta \pm \sqrt{f(s)}\,\bar\xi\)/2$.
The corresponding result for the hyperbolic case $V''/T'' < 0$, e.g.,
close to a quadratic maximum of a cubic potential, is obtained from
Eq.~(\ref{wigpropcubicell}) by replacing $f(s) =
3s^{-2}+1$ and $s = \sinh(\tau/2)$.

%
Equation (\ref{wigpropcubicell}) provides us with a
precise geometric outline that complements the features of
the Wigner propagator pointed out in the sequel of
Eq.~(\ref{wigvleckaction}):

\renewcommand{\theenumi}{\roman{enumi}}
\begin{enumerate}

\item The quantum spot formed by the propagator fills a sector in
phase space of opening angle $\theta = 4\,\arccot\,\sqrt{f(s)}$
(Fig.~\ref{ajedrez}c), rotating by $\tau/2$ if $\tau$ is the angle
coordinate along the classical trajectory. Its nodes form straight
lines running parallel to its sidelines, with separations given by
Airy-function zeros (Fig.~\ref{ajedrez}).

\item For short scaled time $\tau \ll 1$, the oscillatory tail is
quasi-one-dimensional, $\theta \ll 1$, pointing in the negative
$p$-direction if $V''' > 0$ and vice versa, and scales with time as
$\tau^3$. It is periodic in $\tau$ with period $2\pi$ and invariant
under time reversal, $\tau \to \tau_0 - \tau$, $p \to -p$, with
respect to $\tau_0 = 0,\,\pi$.

\item As expected for a uniform approximation,
Eq.~(\ref{wigpropcubicell}) resolves the sharp caustics along the
outlines of the quantum spot present in the van Vleck approximation
(\ref{wigvleckprop}) (Fig.~\ref{ajedrez}b) into a smooth penumbra
(Fig.~\ref{ajedrez}c,d). This restores the correct normalization of
the propagator. On the other hand, it restricts the oscillatory
pattern to straight nodelines while with Eq.~(\ref{wigvleckprop}),
nonlinear deformations owing to higher nonlinearities in the potential
can well be reproduced, see Sec.~\ref{deltaprop}.

\item The finite weight the propagator (\ref{wigpropcubicell})
achieves in the penumbra zone outside the illuminated region of
Eq.~(\ref{wigvleckprop}) can be interpreted as the contribution of
complex trajectories. However, we shall not pursue this issue here.

\end{enumerate}


\subsubsection{Perturbation theory from van Vleck approach}
\label{perturbation}

A uniform approximation equivalent to Eq.~(\ref{wigpropcubicell}) can
be obtained through an alternative route which is particularly helpful
in order to analyze the asymptotics at weak anharmonicity and at short
time and the classical limit for the Wigner propagator: We treat the
weak cubic nonlinearity explicitly as a perturbation of an otherwise
quadratic Hamiltonian
\begin{equation}\label{hampertu}
H(p,q) = \frac{p^2}{2m} + \frac{m\omega^2}{2}q^2 + \epsilon q^3,
\end{equation}
to obtain expressions for the various ingredients of the van Vleck
propagator (\ref{wigvleckprop}) through a perturbation expansion in
$\epsilon$.

From the perturbed orbits ${\bf r}''_{\epsilon}({\bf r}',t)$,
we obtain pairs of trajectories ${\bf r}''_{\epsilon\pm}(t)
= {\bf r}''_{\epsilon}({\bf r}' \pm \check{\bf r}'/2,t)$ with initial
points displaced by $\pm \check{\bf r}'/2$ from ${\bf r}'$, the
corresponding centers $\bar{\bf r}''_{\epsilon}({\bf r}',\check{\bf
r}',t) = ({\bf r}''_{\epsilon +} + {\bf r}''_{\epsilon -})/2$, and the
deviations $\Delta {\bf r}''_{\epsilon}({\bf r}',\check{\bf r}',t)
= \bar{\bf r}''_{\epsilon}({\bf r}',\check{\bf r}',t) - \bar{\bf
r}''_{\epsilon}({\bf r}',{\bf 0},t)$ of these final midpoints from the
endpoint ${\bf r}''_{\epsilon}({\bf r}',t)$ of the perturbed
trajectory starting in ${\bf r}'$,
\begin{equation}\label{deltatrajpert}
\begin{split}
\Delta p''_{\epsilon}({\bf r}',\check{\bf r}',t) &=
\frac{3\epsilon}{8\omega}\,
\check{\bf r}'^{\rm t} \mathsf{F}_p \check{\bf r}'+
\O(\epsilon^2,t^4),\\
\Delta q''_{\epsilon}({\bf r}',\check{\bf r}',t) &=
\frac{3\epsilon}{8m\omega^2}\,
\check{\bf r}'^{\rm t} \mathsf{F}_q \check{\bf r}' +
\O(\epsilon^2,t^5),
\end{split}
\end{equation}
with time-dependent coefficient matrices 
\begin{equation} \label{trajpertmatrix}
\mathsf{F}_p =
\[\begin{array}{cc} \frac{2\tau^3}{3} & \tau^2 \\
\tau^2 & 2\(\tau-\frac{\tau^3}{3}\) \end{array}\], \quad
\mathsf{F}_q =
\[\begin{array}{cc} \frac{\tau^4}{6} & \frac{\tau^3}{3} \\
\frac{\tau^3}{3} & 2\(\tau^2-\frac{\tau^4}{6}\) \end{array}\].
\end{equation}
Arising from a perturbation expansion of the equations of motion,
Eqs.~(\ref{deltatrajpert}, \ref{trajpertmatrix}) are reliable only up
to leading order in $\tau$, hence their validity is also restricted in
time to $\tau \ll 2\pi$. They constitute the classical skeleton of the
quantum spot and provide the input for its determinantal prefactor
and action.

\begin{figure}[h!]
 \includegraphics[scale=0.5]{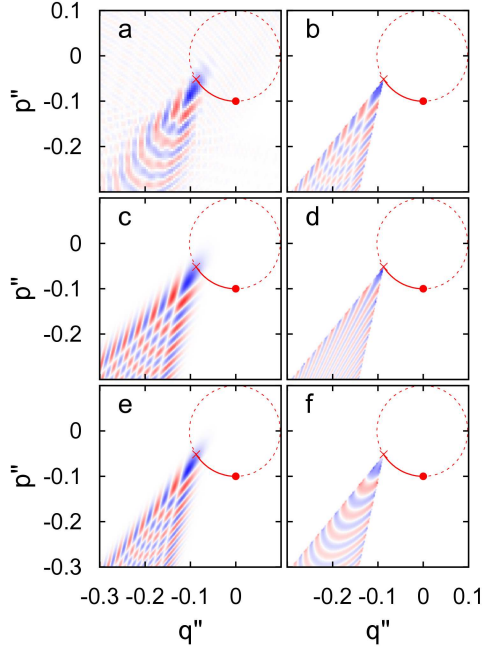}
\caption{\label{ajedrez} 
Different versions of the Wigner propagator for a harmonic potential
with weak cubic anharmonicity as a function of the final phase-space
coordinates $(q'',p'')$ at a scaled time $\omega t = \pi/3$: Wigner
propagator in the semiclassical approximation
(\protect\ref{wigpropcubicell}) based on the exact quantum result
(\protect\ref{wigndqmprop}) (panel a), the van Vleck approximation
(\protect\ref{wigpropvleckpert}) (b), phase-space path integration
(c), and short-time version (\protect\ref{wigpropcubicpert}) of the 
same approach (e). Panels d,f are contributions to
Eq.~(\protect\ref{wigpropvleckpert}) corresponding to hyperbolic
(cosine term, d) and elliptic (sine term, f) trajectory pairs,
respectively, cf.\ Fig.~\protect\ref{zanahoria}b,c. Parameter values
are $\hbar = 0.01$, $\epsilon = 0.2$, $m = 1$, $\omega = 1$. The
underlying classical trajectory (red line in all panels) has been
launched at $(q',p') = (0,-0.1)$. Color code ranges from red
(negative) through white (zero) through blue (positive).
}
\end{figure}

It is, however, more convenient to express these quantities in terms
of $\Delta {\bf r}''_{\epsilon}$, the final argument of the
propagator relative to the classical trajectory starting in ${\bf
r}'$, instead of the initial displacement $\check{\bf r}'$. By means
of a few elementary phase-space transformations that diagonalize
simultaneously the two coefficient matrices (\ref{trajpertmatrix}),
the pair of quadratic equations (\ref{deltatrajpert}) can be resolved
for $\check {\bf r}'$ as a function of $\Delta {\bf
r}''_{\epsilon}$. We thus obtain the propagator in the form
\begin{align}\label{wigpropvleckpert}
& G_{\rm W}^{\rm vV}({\bf r}'',{\bf r}',t) = 
\frac{2\pi\hbar\beta}{\[2\frac{t}{m}\Delta p''^2 -
12\Delta p'' \Delta q'' + 12\frac{m}{t}\Delta q''^2\]^{1/4}}
\notag\\
\times &\left\{\cos\[\beta(\Delta r''_+ - \Delta r''_-)^{3/2}\]
+\sin\[\beta(\Delta r''_+ + \Delta r''_-)^{3/2}\]\right\},
\end{align}
with $\beta = 2m^{3/4}/3^{3/2}\hbar\epsilon^{1/2}t^{5/4}$ and
$\Delta r''_\pm = (1\mp\sqrt{3})\sqrt{t/m}\Delta p'' \pm
2\sqrt{3m/t}\Delta q''$.
We identify the cosine term in Eq.~(\ref{wigpropvleckpert}) as the
contribution of the pair of hyperbolic trajectories (saddles of the
action, Fig.~\ref{ajedrez}e) and the sine term as the elliptic
contribution (extrema, Fig.~\ref{ajedrez}f), cf.\
remark (\ref{ellhypshells}) after Eq.~(\ref{wigvleckaction}) and
Fig.~\ref{zanahoria}c.

Unfortunately, Eq.~(\ref{wigpropvleckpert}) becomes spurious precisely
in the limit $\epsilon \to 0$ for which it has been devised, since in
this limit, the stationary points of the action coalesce and the
stationary-phase approximation underlying Eq.~(\ref{wigpropvleckpert})
breaks down. The problem can be fixed, however, in a straightforward
manner: Since the action in Eq.~(\ref{wigvleck}) is odd in
the integration variable $R$ and the two known trajectory pairs
entering Eq.~(\ref{wigpropvleckpert}) mark its four stationary points
${\bf R}_{--} = -{\bf R}_{++}$ and ${\bf R}_{-+} = -{\bf R}_{+-}$
(Fig.~\ref{zanahoria}c), corresponding resp.\ to the elliptic
and hyperbolic pairs, we can uniquely reconstruct an effective action
that contains only linear and cubic terms in $R$
(Fig.~\ref{zanahoria}c) and thus permits an exact evaluation
of the $R$-integration. The result (cf.\
Fig.~\ref{ajedrez}c), replacing Eq.~(\ref{wigpropvleckpert}), reads
\begin{equation}\label{wigpropcubicpert}
G_{\rm W}^{\rm pi}({\bf r}'',{\bf r}',t) = 
\frac{4\sqrt{3}}{\alpha^{2/3}}
\Ai\(\frac{\Delta r''_-}{\alpha^{1/3}}\)
\Ai\(\frac{\Delta r''_+}{\alpha^{1/3}}\),
\end{equation}
with $\alpha = 3\epsilon\hbar^2 t^{5/2}m^{-3/2}$ and
$\Delta r''_-$, $\Delta r''_+$ as above.
It follows from Eq.~(\ref{wigpropcubicell}) as a short-time
approximation, to leading order in $\tau$. Equation
(\ref{wigpropvleckpert}), in turn, is recovered if we replace the Airy
functions by their asymptotics for large negative argument
\cite{AS84}, $\Ai(-x) \to \pi^{-1/2} (\frac{3}{2})^{1/4} x^{-3/8}
\sin(\frac{2}{3}x^{3/2}+\frac{\pi}{4})$.

Equation (\ref{wigpropcubicpert}) has been obtained combining
perturbation theory with semiclassics on the level of the van Vleck
propagator augmented by a uniform approximation---a route that might
appear less systematic and controlled than the access
through phase-space path integration that led to
Eq.~(\ref{wigpropcubicell}). Its virtue, though, lies
in the fact that it is based on trajectory pairs, hence readily
extends to a higher number of degrees of freedom,
while this is far from obvious in the case of path integration. This
feature together with its simplicity and the fact that is is adapted
to short propagation time suggests Eq.~(\ref{wigpropcubicpert}) as a
suitable choice for on-the-fly molecular-dynamics
applications \cite{MTM99,DM02a}.

\subsubsection{Asymptotics $\epsilon \to 0$, $\hbar \to 0$, $t \to 0$}
\label{asymptotics}
As pointed out the discussion of
Eq.~(\ref{wigvleckaction}), the semiclassical Wigner
propagator replaces the Liouville propagator by a smooth quantum spot
only if the potential is not purely harmonic. It should
converge to the Liouville propagator for vanishing
anharmonicity. Indeed, invoking the asymptotic expression
$\lim_{\kappa\to 0}\kappa^{-1} \Ai(x/\kappa) = \delta(x)$ \cite{VS04}
for the Airy functions in Eq.~(\ref{wigpropcubicell}), we find for
$\alpha \to 0$,
$G_{\rm W}({\bf r}'',{\bf r}',t) = 4\sqrt{3}
\delta(\Delta r''_-)\delta(\Delta r''_+)$.
Separating $\Delta p''$ and $\Delta q''$ in the arguments of the delta
functions, this proves to be equivalent to
$G_{\rm W}({\bf r}'',{\bf r}',t) = \delta(\Delta{\bf r}'')$,
the classical Liouville propagator. Considering the dependence of
$\alpha$ on $\epsilon$, $\hbar$, and $t$, this includes the classical
limit (for subtleties of this limit for Wigner functions and their
propagation, see \cite{CF&08}) and the limit $t \to 0$. Moreover, the
powers $1$,  $2$, and $2.5$, respectively, with which $\epsilon$,
$\hbar$, and $t$ appear in $\alpha$, indicate a hierarchy of the
corresponding limits: At finite $\hbar$, the smooth quantum spot
already collapses to a delta function in the limit of weak
anharmonicity, i.e., for a quantum harmonic oscillator. And even at
finite $\epsilon$, taking into account the different scaling with time
of $\Delta q''$ and $\Delta p''$ as they enter
Eq.~(\ref{wigpropvleckpert}), the lateral ($\bar\xi$) extension of the
quantum spot disappears first with $t \to 0$ while its longitudinal
($\bar\eta$) dimension (in $\Delta p''$-direction) scales with $t$ as
it does with $\epsilon$.

\section{Numerical results}
\label{numerics}

\subsection{Models}
\label{models}

\subsubsection{Morse oscillator}
\label{morse}

We choose the Morse oscillator for a detailed study of semiclassical
Wigner propagation for two reasons: Above all, being prototypical
for strongly anharmonic molecular potentials and correspondingly
complex dynamics \cite{KK04,BB06,BL07,RG09}, it is a widely used
benchmark for numerical methods in this realm
\cite{DS88,SG96,Gro99,HRG04,KRO04}. Secondly, the Morse oscillator has
the remarkable advantage, shared only with very few other potentials,
that closed analytical expressions are available not only for its
energy eigenstates \cite{Mor29} but even for their Wigner
representations \cite{DS88}, a feature that greatly facilitates the
comparison with exact quantum-mechanical results
\cite{LS82,DS88,HM&98a,HM&98bc,FR&00,SK06} and hence
provides a solid basis for an objective test of the performance of
semiclassical approximations to the propagator.

The one-dimensional Morse potential \cite{Mor29}
\begin{equation}\label{onemorse}
V_{\rm Morse}(q) = D(1 - {\rm e}^{-a q})^2
\end{equation}
is determined by the depth $D$ and inverse width $a$ of the potential
well. Quantum-mechanically, its spectrum is discrete for $0 < E < D$
and continuous for $E > D$. The number of bound states
$|\alpha\rangle$ is given (up to $\O(1)$) by the parameter
$\lambda=\sqrt{2mD}/a \hbar$, to be interpreted as an inverse Planck's
constant in natural units. Analytical expressions in
terms of Laguerre polynomials are available for the
eigenfunctions $\psi_{\alpha}(q) = \langle q|\alpha\rangle$, see
Refs.\ \cite{Mor29,DS88}. Applying the Weyl transform (\ref{wigfunc})
to $\hat\rho = |\alpha\rangle\langle\alpha|$ then leads to the
corresponding Wigner eigenfunctions $W_{\alpha\alpha}({\bf r})$
\cite{DS88}.
Exact solutions for the continuum states, on the other hand, are not
known. Numerical evidence indicates, however, that the error caused by
their omission in the basis set underlying quantum calculations of the
time evolution is acceptable as long as the energy does not come too
close to the threshold $E = D$.

\subsubsection{Quartic double well}
\label{quartic}

The quartic double-well potential is in many respects complementary to
the Morse oscillator. From a phenomenological point of view, it is the
standard model for the study of coherent tunneling, and therefore
constitutes a particularly hard problem for semiclassical
propagation. Technically, it is characterized by the absence
of a continuum, which facilitates quantum calculations. At the same
time, no analytical expressions for its eigenfunctions are known.

To begin with, define the quartic double-well potential as
$V(x)=-m\omega^2 x^2/4 + m^2\omega^4 x^4/64 E_{\rm b}$,
where $\omega$ is the oscillation frequency near the minima at $x_\pm
= \pm \sqrt{8E_{\rm b}/m\omega^2}$ and $E_{\rm b}$ their depth. In
natural units $q=\sqrt{m\omega/\hbar}\,x$ and $\tau = \omega t$, the
Schr\"odinger equation reads
\begin{equation}\label{qdwham}
\im\frac{\partial}{\partial t} |\psi\rangle =
\hat H |\psi\rangle, \quad
\hat H(\hat p,\hat q) = \frac{\hat p^2}{2} - \frac{\hat q^2}{4} +
\frac{\hat q^4}{64 \Delta}.
\end{equation}
Its only parameter, the dimensionless barrier height $\Delta=E_{\rm
b}/\hbar\omega$, measures approximately the number of tunneling
doublets, half the number of eigenstates below the barrier top.

Eigenvalues and eigenfunctions underlying the exact quantum
calculations shown in the sequel have been obtained in a basis of
harmonic-oscillator eigenstates, with unit frequency and centered at
$q = 0$.

\subsection{Propagating delta functions: the propagator as a
stand-alone quantity}
\label{deltaprop}

To our best knowledge, no detailed account of the propagator of the
Wigner function as a quantity of its own right has been published to
date, except for the recent Ref.~\cite{DVS06}. Indeed,
it is inaccessible to semiclassical approximation
schemes based on Gaussian smoothing. We therefore find it appropriate
to illustrate its basic properties with data obtained for the two
models featured in this section. Moreover, the interference
pattern characterizing the Wigner propagator is a sensitive diagnostic
for the performance of the different semiclassical
approximations we are proposing. In addition, an analysis of the
propagator allows to assess by mere inspection the validity of
propagation schemes based on non-classical but deterministic
trajectories \cite{LS82,Lee92,Raz96,DM01,THW03}, see
Sec.~\ref{stationary}.

The naked propagator is equivalent to the time evolution of a
delta function as initial Wigner function, which is not an admissible
Hilbert-space element. Notwithstanding, since Wigner functions
are measurable, e.g., through quantum-state tomography \cite{SB&93},
so is the propagator, under weak conditions \cite{fn1}, by
unfolding the final Wigner function from the initial one, which
adds legitimacy to considering the propagator as a stand-alone quantity.

\begin{figure}[h!]
\begin{center}
 \includegraphics[scale=0.3]{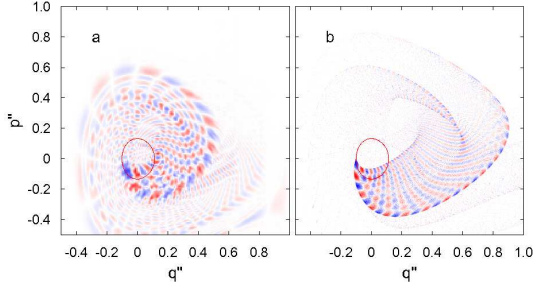}
\caption{Van-Vleck-based semiclassical Wigner propagator
(\protect\ref{wigvleckprop}, \protect\ref{wigvleckaction})
(b) and corresponding exact quantum result
(\protect\ref{wigndqmprop}) (a) as a function of the final phase-space
coordinates $(q'',p'')$ for the Morse oscillator (\protect\ref{onemorse})
at intermediate energy $E = 0.018$ at $t = 10$, approximately the
period of the underlying classical orbit (red line). Parameters are $m
= 0.5$, $D = 1$, $a = 1.25$, $\hbar = 0.005$, the initial point is
$(q',p') = (-0.1,0)$. Color code as in Fig.~\protect\ref{ajedrez}.
}\label{rafaga}
\end{center}
\end{figure}

The linear checkerboard structure of the quantum spot appearing in
Fig.~\ref{ajedrez} is owed to the cubic anharmonicity of the potential
to which the figure refers (its poor resolution in the quantum result,
panel a, is a numerical artefact). In Fig.~\ref{rafaga}, we contrast
it with data for the Morse oscillator, at an intermediate energy $E =
0.018D$  where the nonlinearity is already considerable, and after a
propagation time $\omega t = 10$ of the order of the period of the
underlying classical orbit. The pattern of Fig.~\ref{ajedrez} can
still be discerned but is severely distorted. The nodelines are now
strongly curved, and the pattern even folds over, giving rise to a
phase-space swallow-tail catastrophe. The van-Vleck-based
semiclassical propagator (\ref{wigvleckprop}, \ref{wigvleckaction})
(panel b) captures this complicated structure surprisingly well. The
path-integral-based approximation, by contrast, does not reproduce
these distortions (not shown).

\begin{figure}[h!]
\begin{center}
 \includegraphics[scale=0.3]{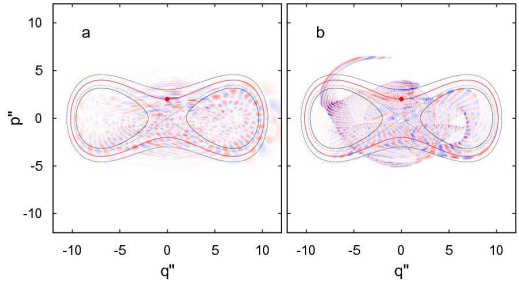}
\caption{As Fig.~\protect\ref{rafaga} but for the
quartic double well (\protect\ref{qdwham}) with $\Delta = 6$ at energy
$E = 2$ above the barrier at $t = 5$, approximately the
period of the underlying classical orbit (red line). The initial point
is $(q',p') = (0,2)$. Color code as in Fig.~\protect\ref{ajedrez}.
}\label{hojarasca}
\end{center}
\end{figure}

Figure \ref{hojarasca} shows a similar comparison for the quartic
double well, at an energy above the barrier top. Here, the
interference pattern is even more complex than in
Fig.~\ref{rafaga}. While the van-Vleck-based propagator does not agree
with the quantum calculation in the intricate fine details of the
oscillatory fringes, it correctly reproduces much of the more global
features of the distribution. For an account of its performance at
energies below the barrier top, see Sec.~\ref{tunneling}.

\subsection{Propagating Gaussians}
\label{gaussians}

Gaussian initial states deserve their ubiquity not only for
their unique physical properties, exemplified by
coherent states. In the context of the Wigner representation,
Gaussians gain special relevance as they constitute the only
admissible Wigner functions that are positive definite and therefore
can be interpreted in probabilistic terms \cite{Hud74}. They have
achieved a fundamental r\^ole for semiclassical propagation as they
provide a natural smoothing which allows to reduce the time evolution
of an entire phase-space region to the propagation along a single
classical trajectory.

By difference to Gaussian-wavepacket propagation and semiclassical
IVRs, however, the semiclassical Wigner propagator does not involve any
smoothing by construction and thus allows to time-evolve
arbitrary initial states, Gaussian or not. A more specific question is
then how semiclassical approximations to it operate
on the space of Gaussian phase-space distributions. We find that
they generally transform them into a broader class of functions
which, in view of the above theorem, can no longer remain positive
definite.

Define Gaussians in phase space \cite{Hel91a} by
\begin{equation}\label{wiggauss}
W({\bf r}) = \frac{\sqrt{\det\mathsf{A}}}{(2\pi\hbar)^f}
\exp\[-\frac{({\bf r}-{\bf r}_0)^{\rm t}\mathsf{A}({\bf r}-{\bf r}_0)}
{2\hbar}\].
\end{equation}
The $2f\times 2f$ covariance matrix $\mathsf{A}$ controls size,
shape, and orientation of the Gaussian centered in ${\bf r}_0 = ({\bf
p}_0,{\bf q}_0)$. The more specific class of minimum-uncertainty
Gaussians, equivalent to Wigner representations of coherent states
\cite{Sch01}, is characterized by $\det\mathsf{A} = 1$. In what
follows, in two dimensions ${\bf r} = (p,q)$, we choose
$\mathsf{A} = \hbox{diag}\,(1/\gamma,\gamma)$, so that
\begin{equation}\label{wigmingauss}
W({\bf r})=\frac{1}{\pi\hbar}
\exp\[-\frac{(p-p_0)^2 + \gamma^2(q-q_0)^2}{\gamma\hbar}\],
\end{equation}
equivalent to a position-space wave function \cite{Hel91a}
$\psi ({\bf q})=(\pi\hbar)^{-f/4} \exp[-\frac{\gamma}{2\hbar}
|q-q_0|^2 + \frac{\im}{\hbar}{\bf p}_0\cdot({\bf q}-{\bf q}_0)]$.

\subsubsection{Autocorrelation functions}
\label{autocorrelation}

A standard interface between dynamical and spectral data,
autocorrelation functions have a wide range of applications in atomic
and molecular physics and provide a robust and easily verifiable assay
of the accuracy and efficiency of propagation methods. The overlap
$C_{\psi}(t) = \langle\psi(t)|\psi(0)\rangle$ for an initial state 
$|\psi(0)\rangle$ upon squaring readily translates into an expression in
terms of the Wigner representation $W_\psi({\bf r})$ of that state,
$|C_{\psi}(t)|^2 = (2\pi\hbar)^f\int\d^{2f} r\,
W_\psi({\bf r},0)W_\psi({\bf r},t)$,
or, involving the propagator explicitly,
\begin{equation} \label{wigpropauto}
|C_{\psi}(t)|^2 = (2\pi\hbar)^f\int\d^{2f} r''\int\d^{2f} r'\,
W_\psi({\bf r}'',0)
G_{\rm W}({\bf r}'',{\bf r}',t) W_\psi({\bf r}',0).
\end{equation}

\begin{figure}[h!]
\begin{center}
 \includegraphics[scale=.32]{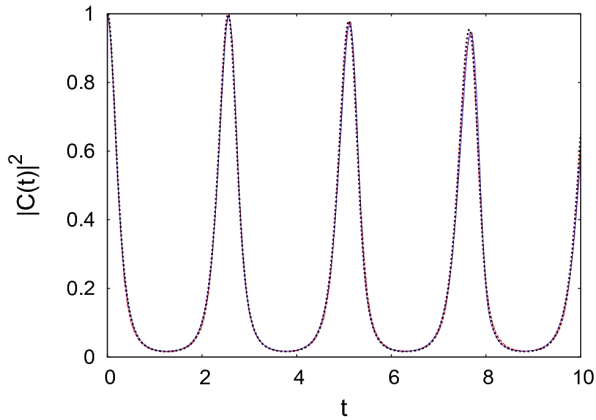}
\caption{Autocorrelation function (\protect\ref{wigpropauto}) for a
Gaussian initial state (\protect\ref{wiggauss}) in a Morse potential
(\protect\ref{onemorse}), for the semiclassical approximation
(\protect\ref{wigvleckprop}, \protect\ref{wigvleckaction}) (dashed
line, red), compared to an exact quantum
(Eq.~(\protect\ref{wigndqmprop}), full line, blue) and a classical
calculation (Eq.~(\protect\ref{liouprop}), dotted, black) of the
Wigner propagator. Parameter values are $m = 0.5$, $D = 1$, $a =
1.25$, $\hbar = 0.005$, the initial centroid is $(q_0,p_0) =
(-0.1,0)$.
}\label{vaivenmorse}
\end{center}
\end{figure}

%
\begin{figure}[h!]
\begin{center}
 \includegraphics[scale=0.32]{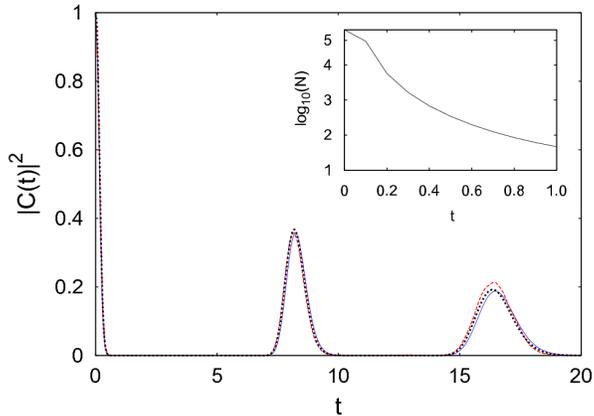}
\caption{As Fig.~\protect\ref{vaivenmorse} but for a quartic double
well (\protect\ref{qdwham}) with $\Delta = 6$ at an energy $E =
1.0026$ above the barrier top. The initial centroid of the Gaussian is
$(q_0,p_0) = (4,7)$. Inset: Nonconservation of the norm $N$ for the
same propagation process but without interspersed normalization steps.
}\label{vaivenqdw}
\end{center}
\end{figure}

In Fig.~\ref{vaivenmorse}, we compare the autocorrelation functions
(\ref{wigpropauto}) for the Morse oscillator with a Gaussian initial
state (\ref{wiggauss}) as obtained by an exact quantum calculation
(full line) to the semiclassical approximation (\ref{wigvleckprop},
\ref{wigvleckaction}) (dashed) and the classical propagator
(\ref{liouprop}) (dotted) at an intermediate energy. Over the roughly
four periods of the classical trajectory monitored, the agreement is
impressive. Figure \ref{vaivenqdw} is an analogous comparison but for
the quartic double-well potential. Here, the discrepancy between
semiclassical and quantum data is slightly more significant.

In order to obtain the data underlying Figs.~\ref{vaivenmorse},
\ref{vaivenqdw}, it was necessary to periodically renormalize the
propagator, compensating for a lack of norm conservation as is
notorious for van-Vleck-based approximations
\cite{Kay94}. In Fig.~\ref{vaivenqdw} (inset), we are monitoring the
loss of the norm of the final Wigner function for a typical run of the
propagator (\ref{wigvleckprop}, \ref{wigvleckaction}) without
renormalization in the quartic double well over a fraction of the
period of the underlying classical orbit.

\subsubsection{Resolving phase-space structures}
\label{phasespace}

More detailed information on the performance of propagators and
reasons for their possible failure than is contained in
autocorrelation functions can be extracted from the full phase-space
structures of propagated Gaussians. In particular in strongly
nonlinear systems, the degree to which a semiclassical propagator is
able to reproduce deviations of the time-evolved states from Gaussians
is a sensitive measure of its quality.

\begin{figure}[h!]
\begin{center}
 \includegraphics[scale=0.5]{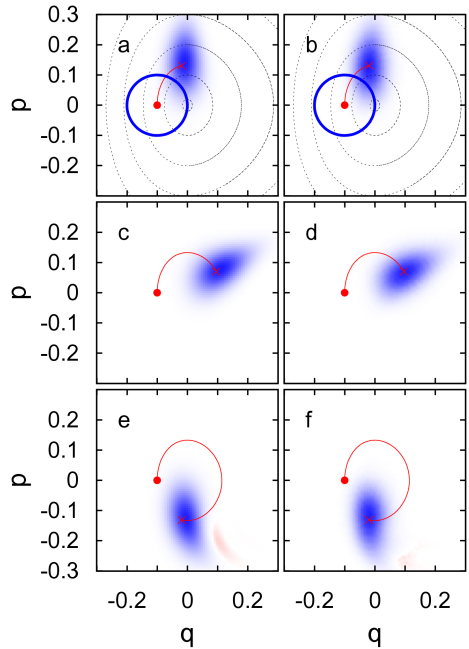}
\caption{Time evolution of a state prepared as a
minimum-uncertainty Gaussian (\protect\ref{wigmingauss}) (bold blue
circle in panels a,b is its contour enclosing a Planck cell) in the
Morse potential (\protect\ref{onemorse}) (black contour lines in panel a)
for the semiclassical approximation (\protect\ref{wigvleckprop},
\protect\ref{wigvleckaction}) (right column), compared to an exact
quantum calculation (\protect\ref{wigndqmprop}) (left), at
times $t = 0.5072$ (panels a,b), 1.0144 (c,d), 2.0288 (e,f). The full
red line is the classical orbit of the initial centroid $(q_0,p_0) =
(-0.1,0)$. Parameter values are $m = 0.5$, $D = 1$, $a = 1.25$, $\hbar
= 0.005$. Color code as in Fig.~\protect\ref{ajedrez}.
}\label{plumitamorse}
\end{center}
\end{figure}

\begin{figure}[h!]
\begin{center}
 \includegraphics[scale=0.3]{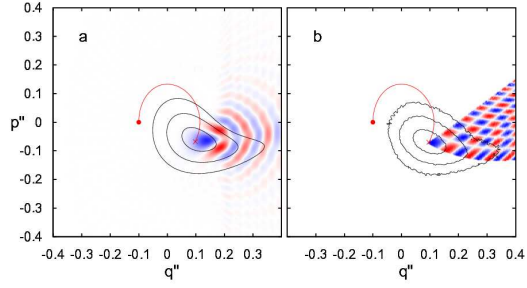}
\caption{Same as Figs.~\protect\ref{plumitamorse}c,d but
comparing the time-evolved Gaussian (contour lines) to
the respective propagator (\protect\ref{wigvleckprop},
\protect\ref{wigvleckaction}) on the centroid orbit (red) at the same
time $t = 1.0144$. Color code as in Fig.~\protect\ref{ajedrez}.
}\label{plumamorse}
\end{center}
\end{figure}

\begin{figure}[h!]
\begin{center}
\includegraphics[scale=0.5]{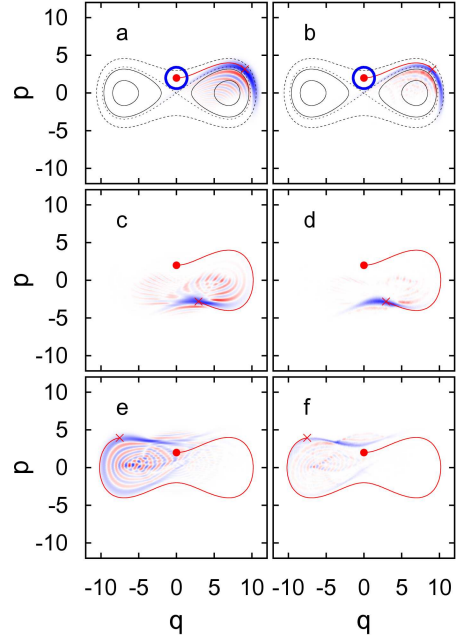}
\caption{As Fig.~\protect\ref{plumitamorse} but for the
quartic double well (\protect\ref{qdwham}) with $\Delta = 6$ at times
$t = 3$ (panels a,b), 9 (c,d), 15 (e,f). The initial centroid is
$(q_0,p_0) = (0,2)$. Color code as in Fig.~\protect\ref{ajedrez}.
}\label{plumitaqdw}
\end{center}
\end{figure}

Such deviations are unmistakable in Fig.~\ref{plumitamorse},
a series of snapshots of the time evolution of an initially
Gaussian state under the same conditions as for the autocorrelation
function depicted in Fig.~\ref{vaivenmorse}. The asymmetric droplet
shape of the distribution, notably in panels c to f, is not compatible
with a linearly deformed Gaussian, but is immediately explained by the
form of the underlying propagator on the centroid trajectory,
superimposed on the evolved distribution at the same time
(Fig.~\ref{plumamorse}). The non-elliptic shape is therefore a
consequence of the strongly anharmonic potential. In
Fig.~\ref{plumitamorse}, panels e,f, we even observe
fringes where the distribution takes on negative values. Unavoidable
for non-Gaussian Wigner functions \cite{Hud74}, this behavior appears
even more conspicuous in Fig.~\ref{plumitaqdw}, analogous to
Fig.~\ref{plumitamorse} but for a quartic double well potential. At
the same time, the resolution of these structures is relatively poor,
owing to the reduced density of classical trajectories used in the
calculation towards the periphery of the initial distribution.


A more quantitative account of this feature is obtained considering
the semiclassical propagator (\ref{wigpropcubicell}). Folding
Airy functions with initial minimum-uncertainty Gaussians leads into a
wider function space of polynomials of up to cubic order in the
exponent and, in semiclassical Wigner propagation, replaces the
(frozen, thawed, etc.)\ Gaussians constituting the framework of
Gaussian-wavepacket propagation \cite{Hel75,HK84,Gro99,BA&01,Hel91b}. 
It ranges from strongly distorted, oscillatory distributions
in the ``deep quantum regime'' to near Gaussians in the semiclassical
limit.
%

\subsection{Propagating stationary states}
\label{stationary}

Propagating energy eigenstates of a quantum system may appear a
trivial task. For the propagation of Wigner functions, however, there
is more to gain than just a check of norm conservation, as explained
in a remarkable work by Lee and Scully \cite{LS82}. They argue that,
while for a system with anharmonic potential a phase-space flow along
classical trajectories cannot reproduce the quantum dynamics, at
least in the case of energy eigenstates they may be
replaced by another type of characteristic: Evidently, Wigner
functions associated to eigenstates are invariant under a flow that
follows their contour lines, suggesting the latter as a kind of
``Wigner trajectories'' to replace the classical ones for finite
$\hbar$.

In the present subsection, we confront their proposal
with semiclassical Wigner propagation along pairs of
classical trajectories, taking advantage of the fact that also
Ref.~\cite{LS82} is based on a study of the Morse oscillator. We not
only find that our method describes the quantum evolution of
eigenstates to high accuracy in terms of autocorrelations (far better
than \cite{LS82}) but even present convincing arguments why the very
concept of a quantum phase-space flow along ``Wigner trajectories'' is
misleading.

\begin{figure}[h!]
\begin{center}
 \includegraphics[scale=0.3]{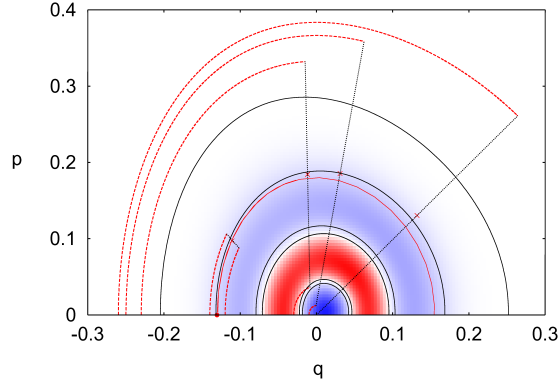}
\end{center}
\caption{Wigner representation
of the second excited state of the Morse oscillator \cite{DS88} (black
contour lines and color code as in Fig.~\protect\ref{ajedrez}),
compared to the classical orbit ${\bf r}_{E_2}^{\rm cl}(t)$ at the
corresponding eigenenergy $E_2$ (full red line) and to midpoint paths
$\bar{\bf  r}_j(t) = [{\bf r}_{j_-}^{\rm cl}(t) + {\bf r}_{j_+}^{\rm
cl}(t)]/2$ for pairs of classical trajectories ${\bf r}_{j_\pm}^{\rm
cl}(t)$ (dashed red lines), with common initial midpoint $\bar{\bf r}'
= {\bf r}_{E_2}^{\rm cl}(0)$ but increasing initial separation
$\check{\bf r}'_j \equiv (\check p'_j,\check q'_j) = {\bf r}'_{j_+} -
{\bf r}'_{j_-}$ with $\check p'_j = 0$ and $\check q'_j = 0.002$,
$0.2$, $0.24$, $0.26$ (see text). Parameter values are $m = 0.5$, $D =
1$, $a = 1.25$, $\hbar = 0.005$.
}\label{WTplot1}
\end{figure}

%
In Fig.~\ref{WTplot1} we contrast the contours of the
second excited eigenstate of the Morse oscillator in Wigner
representation \cite{DS88}
with classical
trajectories for the same system. At this point, it is not even clear
which contour to compare with which trajectory. As a reasonable choice
which becomes compelling in the classical limit, for a given
eigenstate $|\alpha\rangle$ we focus on the classical trajectory
at $E = E_\alpha$. Along this ``quantizing'' orbit, the corresponding
Wigner eigenfunctions exhibit a narrow ridge, to be chosen as the
relevant contour. We find that the classical trajectory roughly
follows the contour line but deviates in quantitative detail.

We would expect semiclassical approximations to improve systematically
on mere classical propagation. Figure \ref{WTplot1} illustrates
how this is achieved by propagating along midpoints of non-identical
classical trajectory pairs: Keeping the initial point ${\bf r}'$ on a
given Wigner contour fixed, we increase the initial separation
$\check{\bf r}'_j = {\bf r}'_{j_+} - {\bf r}'_{j_-}$ of the trajectory
pairs launched in ${\bf r}'_{j_\pm}$. While for $\check{\bf r}' = 0$
(classical trajectory) the deviation from the contour line is large,
the midpoint paths $\bar{\bf r}_j(t) = [{\bf r}_{j_-}(t) + {\bf
r}'_{j_+}(t)]/2$ indeed appear to move towards the Wigner contour with
increasing $\check{\bf r}'$. However, they do not approach it
asymptotically but continue shifting further past the Wigner contour,
indicating that it plays no particular r\^ole for quantum time
evolution in phase space, not even of eigenstates.

\begin{figure}[h]
\includegraphics[scale=0.3]{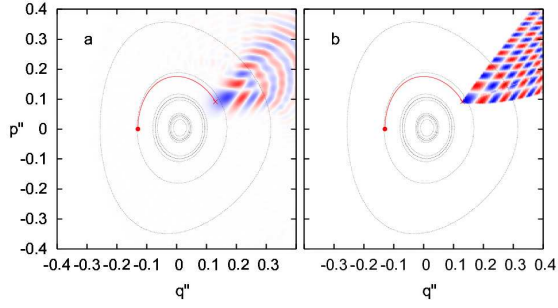}
\caption{Van-Vleck based Wigner propagator
(\protect\ref{wigvleckprop}, \protect\ref{wigvleckaction}) (panel b)
and corresponding exact quantum result (\protect\ref{wigndqmprop}) (a)
for the Morse potential (\protect\ref{onemorse}) as a function of the
final phase-space coordinates $(q'',p'')$, at $t = 1.5216$. The initial
point ${\bf r}' = {\bf r}_{E_2}^{\rm cl}(0)$ is the same as in
Fig.~(\protect\ref{WTplot1}). They are compared to the classical orbit
${\bf r}_{E_2}^{\rm cl}(t)$ (bold line) and the contour of the Wigner
eigenfunction $W_{22}({\bf r})$ (grey), cf.\ Ref.~\cite{DS88},
passing through ${\bf
r}_{E_2}^{\rm cl}(0)$. Parameter values are $m = 0.5$, $D = 1$, $a =
1.25$, $\hbar = 0.005$. Color code as in Fig.~\protect\ref{ajedrez}.
}\label{morsespot}
\end{figure}

In fact, semiclassical Wigner propagation generally does not reduce
to a deterministic phase-space flow, not along classical nor along
modified (quantum) trajectories. In Fig.~\ref{morsespot} we show the
quantum spot formed by the Wigner propagator for the same initial
conditions as in Fig.~\ref{WTplot1}. As in similar plots
throughout this paper, it appears as a smooth distribution that cannot
be replaced by a delta function, neither on the classical trajectory
\textit{nor on the Wigner contour}. It is perhaps surprising but by no
means inexplicable that this propagator, while reshuffling the Wigner
distribution in phase space, nonetheless keeps the Wigner
eigenfunction invariant.

\begin{figure}[h!]
\begin{center}
\includegraphics[scale=0.3]{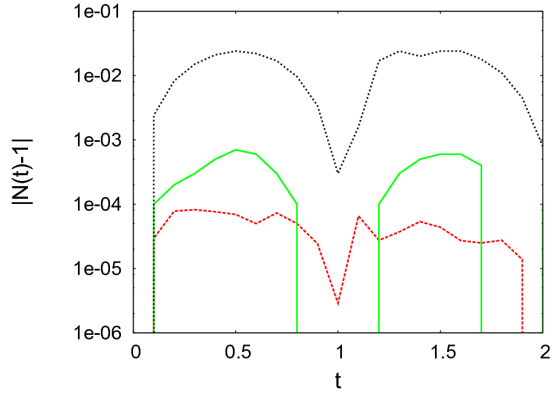}
\end{center}
\caption{Accuracy of the autocorrelation function for the ground state
$W_0({\bf r})$ of the Morse oscillator (equivalent to norm
conservation) propagated with the semiclassical approximation
(\protect\ref{wigvleckprop}, \protect\ref{wigvleckaction}) to the
Wigner propagator (dashed line, red), a deterministic flow along
Wigner contours (full line, green), and the classical Liouville
propagator (\protect\ref{liouprop}) (dotted, black).
Parameter values are $m = 0.5$, $D = 0.15$, $a = 1$, $\hbar = 1.0$.
}\label{WTfidelidad}
\end{figure}

\begin{figure}[ht]
 \includegraphics[scale=0.3]{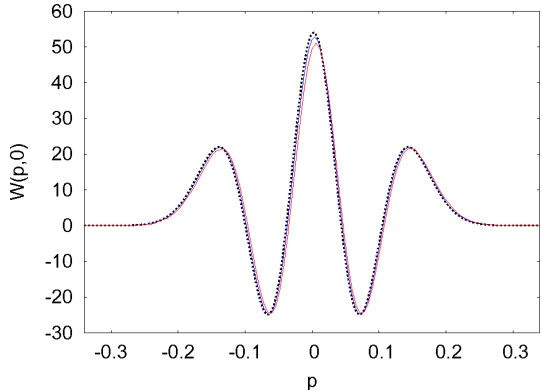}
\caption{Cross section at $p = 0$ of the Wigner eigenfunction
$W_{22}({\bf r})$, cf.\ Ref.~\cite{DS88},
before
(full line, blue) and after propagation through a time $t = 10$
with the semiclassical approximation (\protect\ref{wigvleckprop},
\protect\ref{wigvleckaction}) to the Wigner propagator (dashed line,
red) and the classical Liouville propagator (\protect\ref{liouprop})
(dotted, black).
}\label{WTplot2}
\end{figure}

This is confirmed by Fig.~\ref{WTfidelidad}. We are
plotting the autocorrelation function for the eigenstate
$W_{00}({\bf r})$ as initial state, which in this case coincides with
the norm, propagated with the semiclassical Wigner propagator
(\ref{wigvleckprop}, \ref{wigvleckaction}), and find deviations of the
order of $10^{-5}$ only, as compared to the range of $10^{-4}$ for
propagation along Wigner contours and $10^{-3}$ for mere classical
Wigner dynamics. Moreover, in Fig.~\ref{WTplot2} we compare the
semiclassically propagated state at $t = 10$ with the initial
state. There is no visible difference. We conclude that even in the
case of propagating Wigner eigenstates, an acceptable semiclassical
approximation to the Wigner propagator must not only propagate along
non-classical paths but differ significantly from a delta function on
whatever single trajectory, classical or ``quantum''.


\subsection{Propagating Schr{\"o}dinger cat states}
\label{schroedinger}

The propagation of Sch\"odinger cat states \cite{Zur01} is certainly
not a standard task of molecular dynamics. Yet it is relevant for the
field in various respects: Schr\"odinger cats are a paradigm of
quantum coherence and embody the essence of entanglement in a simple
setting. They allow us to test the performance of propagation methods
in this particular respect in an objective manner, as the separation
of the superposed alternatives and thus the wavelength of the
corresponding interference pattern can be precisely controlled. In
fact, the classic double-slit experiment referred to in
Ref.~\cite{Hel76} to elucidate the challenge quantum coherence poses
to semiclassical propagation is just a contemporary embodiment
of a Schr\"odinger cat, albeit before this term became
popular. Finally, a timely, if fancy, application of this notion are
implementations of quantum computation in medium-size molecules
\cite{VS&01}.

\begin{figure}[h!]
\includegraphics[scale=0.3]{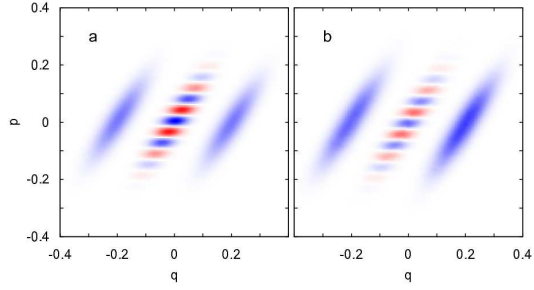}
\caption{Schr\"odinger-cat states time-evolved (\protect\ref{wigcat})
in the Morse potential (\protect\ref{onemorse}) at $t = 0.3$, for
propagation with the semiclassical approximation
(\protect\ref{wigvleckprop}, \protect\ref{wigvleckaction}) to the
Wigner propagator (panel b) as compared to an exact quantum
calculation (\protect\ref{wigndqmprop}) (a). Parameter values are $m =
0.5$, $D = 1$, $a = 0.25$, $\hbar = 0.005$. The initial midpoint and
separation, resp., of the Schr\"odinger cat are $(q_0,p_0) = (0.3,0)$,
$d = 0.2$. Color code as in Fig.~\protect\ref{ajedrez}.
}\label{gata}
\end{figure}

We here consider the propagation of Schr\"odinger cats prepared as the
superposition of two coherent states,
cf.~Eq.~(\ref{wigmingauss}), with variable separation $d$ initially in
position. In the Wigner representation, this amounts to
\begin{equation}\label{wigcat}
W_{\rm cat}({\bf r}) =
W_-({\bf r}) + W_+({\bf r}) + W_{\times}({\bf r}),
\end{equation}
where $W_\pm({\bf r}) = \exp\{-[p_\pm^2 + \gamma^2
q_\pm^2]/\gamma\hbar\}/\pi$, ${\bf r}_\pm = {\bf r} - [{\bf r}_0 \pm
(0,d)]$, while 
$W_{\times}({\bf r}) = \exp\{-[(p-p_0)^2 + \gamma^2(q-q_0)^2]/\gamma\hbar\}
\times\cos[2(p-p_0)d/\hbar]$
encodes the quantum coherence in terms of ``sub-Planckian''
oscillations of wavelength $\hbar/d$ in $p$ \cite{Zur01}. This initial
state is propagated with the semiclassical propagator
(\ref{wigvleckprop}, \ref{wigvleckaction}) in the Morse potential from
the same initial position ${\bf r}_0$ as in
Fig.~\ref{plumitamorse}. The result is compared in Fig.~\ref{gata} to
the exact quantum calculation. Apart from minor deviations in the shape
of the Gaussian envelope, the interference pattern is faithfully
reproduced. This is not surprising in view of the trajectory-pair
construction underlying our semiclassical approximation:

It is instructive to see why propagating along the two
classical trajectories of the respective centroids of the two
``classical'' Gaussians $W_\pm({\bf r})$ already reproduces
essentially the sub-Planckian oscillations. The propagator launched
from the centroid ${\bf r}_0$ of $W_{\times}({\bf r})$
then comprises two terms, $G_{\rm W}({\bf r}'',{\bf r}_0,t) =
G_{{\rm W}0}({\bf r}'',{\bf r}_0,t) + G_{{\rm W}\times}({\bf r}'',{\bf
r}_0,t)$. According to Eqs.~(\ref{wigvleckprop}, \ref{wigvleckaction}),
the first one, propagating along ${\bf r}^{\rm cl}({\bf r}_0,t)$,
bears no oscillating phase factor and therefore practically
cancels upon convolution with the strongly oscillatory
$W_{\times}({\bf r}')$. The second one, by contrast, is the
contribution of the two centroid orbits ${\bf r}^{\rm cl}({\bf
r}_\pm,t)$ forming a pair of non-identical trajectories. It travels
along the non-classical midpoint path $\bar{\bf r}_\times(t) = ({\bf
r}^{\rm cl}({\bf r}_-,t) + {\bf r}^{\rm cl}({\bf r}_+,t))/2$ and
carries a factor $\sim\cos[(2/\hbar)({\bf r}_+ - {\bf r}_-)\wedge{\bf
r}'] = \cos[2dp'/\hbar]$ which couples resonantly to the oscillations
in $W_{\times}({\bf r}')$.

\subsection{Tunneling}
\label{tunneling}

Tunneling is to be regarded a quantum coherence effect ``of
infinite order in $\hbar$'' \cite{LL65}. One therefore does not expect
a particularly good performance of semiclassical methods
in its description, despite various efforts that have been made to
improve them in this respect. Above all, the complexification of phase
space provides a systematic approach to include tunneling in a
semiclassical framework \cite{BM72,Cre94,XA97}.
Here, by contrast, we restrict ourselves to real phase
space, in order not to loose the valuable close relationship between
Wigner and classical dynamics. Even so, we expect that in this
framework tunneling can be reproduced to a certain degree
\cite{BV90,Nov05}. To be sure, Wigner dynamics (in real phase space) is
exact for harmonic potentials. This includes parabolic barriers and
hence a specific case of tunneling, as
pointed out and explained by Balazs and Voros \cite{BV90}: Since
the Wigner propagator invariably follows classical trajectories, the
explanation rather refers to the initial condition in Wigner
representation: Owing to quantum uncertainty, it spills over the
separatrix even if it is concentrated at negative energies, and thus
is transported in part with the classical flow to the other side
of the barrier.

%
\begin{figure}[h!]
\begin{center}
\includegraphics[scale=0.3]{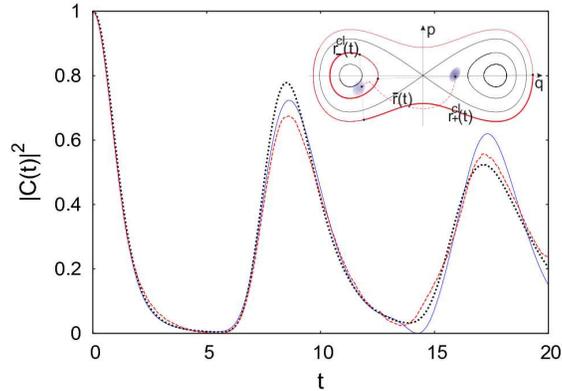}
\caption{Autocorrelation function (\protect\ref{wigpropauto}) for a
Gaussian initial state (\protect\ref{wiggauss}) in a quartic double
well (\protect\ref{qdwham}) with $\Delta = 6$ at an energy $E =
-0.958333$ below the barrier top, for the semiclassical approximation
(\protect\ref{wigvleckprop}, \protect\ref{wigvleckaction}) (dashed
line, red), compared to an exact quantum
(Eq.~(\protect\ref{wigndqmprop}), full line, blue) and a classical
calculation (Eq.~(\protect\ref{liouprop}, dotted, black) of the Wigner
propagator. The initial centroid of the Gaussian is $(q_0,p_0) =
(2,0)$. Inset: Semiclassical description of coherent tunneling in
terms of trajectory pairs, in the framework of the van-Vleck based
Wigner propagator (\protect\ref{wigvleckprop},
\protect\ref{wigvleckaction}). A wavepacket initially prepared near
the right minimum of a double-well potential (blue patch) is
transported along a non-classical midpoint path $\bar{\bf r}(t) = ({\bf
r}_-^{\rm cl}(t) + {\bf r}_+^{\rm cl}(t))/2$
(dashed red line) into the opposite well if the two classical orbits
${\bf r}_\pm^{\rm cl}(t)$ (full red lines) underlying this path
are sufficiently separated initially, e.g., ${\bf r}_+^{\rm
cl}$ on the same side but above the barrier, ${\bf r}_-^{\rm cl}$
within the opposite well. Black lines indicate other contours of the
potential and the separatrix.
}\label{vaivenqdwtun}
\end{center}
\end{figure}

\begin{figure}[h!]
\begin{center}
\hspace{-.8cm}
\includegraphics[scale=0.3]{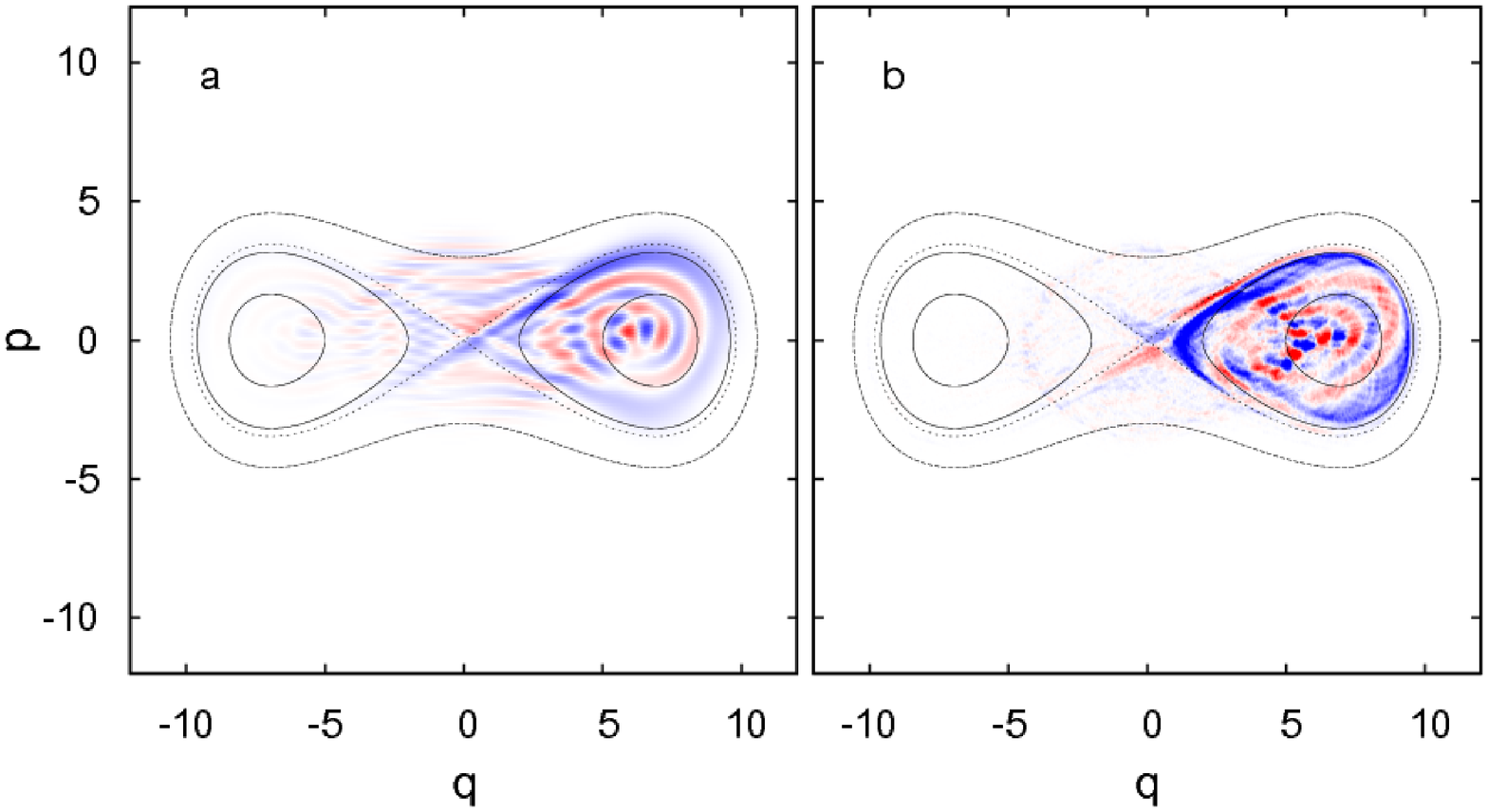}
\caption{Time-evolved state, initially prepared as a
minimum-uncertainty Gaussian (\protect\ref{wigmingauss}) (bold blue
circle in panels a,b is its contour enclosing a Planck cell), at time
$t = 20$ in the quartic double well (\protect\ref{qdwham}) with
$\Delta = 6$ at an energy $E = -0.958333$ below the barrier top, for
the semiclassical approximation (\protect\ref{wigvleckprop},
\protect\ref{wigvleckaction}) (panel b), compared to an exact quantum
calculation (\protect\ref{wigndqmprop}) (a). The initial centroid
is $(q_0,p_0) = (2,0)$. Color code as in Fig.~\protect\ref{ajedrez}.
}\label{plumitaqdwtun}
\end{center}
\end{figure}

This is to be considered as a fortunate exception, though. Other,
more typical cases involving genuine quantum effects, as in
particular coherent tunneling between bound states as we are
considering it here, are not so readily accessible to semiclassical
Wigner dynamics. Quantum tunneling in the Wigner representation,
specifically for localized scattering potentials, has been studied at
depth in \cite{MS96,MS97b}, however without indicating a promising
perspective for semiclassical approximations. We are in a slightly
more favorite situation as the concept of propagation along
trajectory pairs provides a viable option how to reproduce tunneling
by means of a semiclassical Wigner propagator: As illustrated in
Fig.~\ref{vaivenqdwtun} (inset), it is trajectory pairs with
sufficiently separated initial points, probing regions in phase space
classically inaccessible to one another, which lead to transport along
classically forbidden paths.

Figure \ref{vaivenqdwtun} is the autocorrelation function for a
Gaussian initial state prepared at an energy slightly below the
barrier top of the quartic double well (\ref{qdwham}). The
semiclassical Wigner propagator (\protect\ref{wigvleckprop},
\protect\ref{wigvleckaction}) reproduces the revivals but
exaggerates their amplitude. This
significant deviation surprises as it \textit{over}estimates quantum
effects. The reasons become clearer upon analyzing the full
phase-space distribution for the propagated state,
Fig.~\ref{plumitaqdwtun}, at a time corresponding to the right edge of
Fig.~\ref{vaivenqdwtun}. In terms of global features of the
distribution, the agreement of the semiclassical (left) with the
quantum result (right) is good, remarkably even in phase-space regions
classically inaccessible from the initial distribution. However, while
the fine fringes of the quantum distribution reaching out into the
opposite well do appear in the semiclassical result, their relative
weight is not correctly reproduced. This is partially explained by the
incorrect balance between central peak and oscillatory tail in the
semiclassical propagator. Moreover, with $\Delta = 6$, we are in a
marginally semiclassical regime and cannot expect an optimal
performance of the van-Vleck-based Wigner propagator.

\subsection{Propagation in the presence of classical chaos}
\label{chaos}
\begin{figure}[h!]
\begin{center}
\includegraphics[scale=0.3]{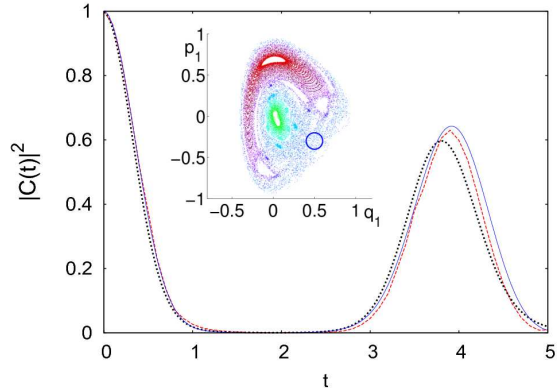}
\caption{As Fig.~\protect\ref{vaivenmorse} but for two coupled
Morse potentials (\protect\ref{twomorse}) at $c = 0.3$. Parameter
values are $m = 1$, $D = 1$, $a = 1.25$, $\hbar = 0.125$, the
initial centroid is $(q_{10},p_{10},q_{20},p_{20}) =
(0.5,-0.3,0.4322,0)$, corresponding to a total energy $E =
0.5$. Inset: Poincar\'e surface of section at $p_2 = 0$ of the
corresponding classical dynamics. Different initial conditions encoded
by colors. The blue circle is the contour of the initial Gaussian
enclosing a Planck cell.
}\label{dosmorse}
\end{center}
\end{figure}

Besides dynamical coherence effects, complex classical dynamics
constitutes a major challenge for propagation schemes in molecular
physics. The two one-dimensional models discussed in the preceding
subsections are strongly anharmonic but remain integrable. In order to
test our method also in the presence of chaos and to check its
scalability towards higher dimensions, we consider a two-freedom
system consisting of Morse oscillators coupled linearly through the
positions, a standard model for complex molecular dynamics
\cite{JZ&01}, with a potential
\begin{equation}\label{twomorse}
V_{\rm 2Morse}(q_1,q_2) =
V_{\rm Morse}(q_1) + V_{\rm Morse}(q_2) + c q_1 q_2,
\end{equation}
where $V_{\rm Morse}(q_i)$, $i = 1,2$ are Morse potentials
(\ref{onemorse}) and $c$ is the coupling parameter. Starting from
regular dynamics at $c = 0$, the system follows the
Kolmogorov-Arnol'd-Moser scenario \cite{LL92} with increasing $c$ and
becomes fully chaotic for $c \gg 1$. We here concentrate on the value
$c = 0.3$, where phase space is mixed. As initial condition, we choose
a two-dimensional minimum-uncertainty Gaussian $W({\bf r}) = W({\bf
r}_1)W({\bf r}_2)$, with $W({\bf r}_i)$, $i = 1,2$, as in
Eq.~(\ref{wigmingauss}), centered within a major chaotic subregion,
see inset of Fig.~\ref{dosmorse}.

The autocorrelation function depicted in the main part of
Fig.~\ref{dosmorse} includes the first major revival. The
van-Vleck-based Wigner propagator reproduces the exact quantum result
reasonably well and shows a tendency to improve on the classical data,
evidencing the r\^ole of dynamical quantum effects in this system.

\section{Algorithms}
\label{algorithms}

In view of the objective to demonstrate the viability of semiclassical
Wigner propagation for numerical applications, we indicate in this
section how to construct suitable algorithms for this purpose, without
entering into details of their implementation. 

\subsection{Assembling the propagator}
\label{props}

\subsubsection{Exact quantum calculation}
\label{qmprop}

For purposes of comparison, calibration, etc.\ it is convenient to
have an algorithm available for a direct exact calculation of the
quantum propagator, independent of semiclassical approximations. We
briefly sketch in the sequel how this is achieved for the propagator
of the Wigner function.

Substituting the unitary time-evolution operator $\hat U(t) =
\sum_\alpha \e^{-\im E_\alpha t/\hbar} |\alpha\rangle\langle\alpha|$
expanded in eigenstates $|\alpha\rangle$, $\hat H|\alpha\rangle =
E_\alpha|\alpha\rangle$, in the definition (\ref{weylprop}), we obtain
the Weyl propagator
$U_{\rm W}({\bf r},t) = \sum_\alpha \e^{-\im E_\alpha t/\hbar}
W_\alpha({\bf r})$ in terms of Wigner eigenfunctions $W_\alpha({\bf
r})$, Eq.~(\ref{wigfunc}) with $\hat\rho = |\alpha\rangle
\langle\alpha|$.
By means of the convolution (\ref{wigweyl}), this yields the Wigner
propagator in the form
\begin{equation}
G_{\rm W}({\bf r}'',{\bf r}',t) = \sum_{\alpha,\beta}
\e^{-\im (E_\beta - E_\alpha) t/\hbar} \int \d^{2f}R\,
\e^{{-\im\/\hbar}({\bf r}''-{\bf r}')\wedge{\bf R}}
\label{wigdiaqmprop}
W_{\alpha\alpha}\(\frac{{\bf r}'+{\bf r}''-{\bf R}}{2}\)
W_{\beta\beta}\(\frac{{\bf r}'+{\bf r}''+{\bf R}}{2}\).
\end{equation}

The convolution in Eq.~(\ref{wigdiaqmprop}) is an
inconvenient feature and can be avoided. Combining it with the
integrations implicit in the definitions of the Wigner eigenfunctions,
it can be evaluated, leading to an expression \cite{AD05}
\begin{equation}\label{wigndqmprop}
G_{\rm W}({\bf r}'',{\bf r}',t) = (2\pi\hbar)^f \sum_{\alpha,\beta}
\e^{\frac{-\im}{\hbar} (E_\beta - E_\alpha) t}
W_{\alpha\beta}^*({\bf r}')W_{\alpha\beta}({\bf r}''),
\end{equation}
which involves generalized Wigner functions \cite{DS88}
\begin{equation}\label{wigndfunc}
W_{\alpha\beta}({\bf r}) = \frac{1}{2\pi\hbar} \int \d^f q' 
\e^{\frac{-\im}{\hbar}{\bf p\cdot q'}} \<{\bf q} +
\frac{{\bf q}'}{2}\right|
\hat\rho_{\alpha\beta}\left|{\bf q} - \frac{{\bf q}'}{2}\>,
\end{equation}
with $\hat\rho_{\alpha\beta} = |\alpha\rangle\langle\beta|$. In
the diagonal case $\alpha = \beta$, they correspond to the Wigner
representation of pure states. For $\alpha \neq \beta$, they are not
generally real as are the diagonal ones (\ref{wigfunc}) but
Hermitian, $W_{\alpha\beta}({\bf r}) = W_{\beta\alpha}^*({\bf
r})$. Both forms, (\ref{wigdiaqmprop}) as well as (\ref{wigndqmprop}),
are consistent with the group properties of the propagator, in
particular with the initial condition $G_{\rm W}({\bf r}'',{\bf r}',0)
= \delta({\bf r}'' - {\bf r}')$. The quantization required to obtain
the basis states $|\alpha\rangle$ can be circumvented by integrating
the Schr\"odinger equation directly (e.g., by split-operator methods)
to propagate the density operator or the wavefunction (for pure
states) in a suitable representation, followed by a Weyl transform
(\ref{wigfunc}).

\subsubsection{Semiclassical approximation based on path integrals} 
\label{piprop}

Equations (\ref{wigpropfourier}, \ref{sigma}) reduce the calculation
of the Wigner propagator in the path-integral-based approximations to
quadratures. More problematic from a practical point of view
is obtaining the necessary input, since it comprises not only
second- but even third-order derivatives, see Eq.~(\ref{sigma}), of
the potential---an unavoidable feature, in view of the
fundamental r\^ole the anharmonicity of the potential plays for Wigner
propagation.

This disadvantage is compensated for by the fact that this
approximation, at least for weak anharmonicity, is accurate
enough to allow for a propagation based on a centroid trajectory only,
in analogy to Gaussian wavepacket propagation. It then suffices to
evaluate representative values $\omega$ and $\sigma$, respectively, of
the second and third derivatives of the potential in the relevant
phase-space region, keeping them constant for an entire propagation
step, Eqs.~(\ref{wigpropcubicell}) or (\ref{wigpropcubicpert}),
till the next update, e.g., in on-the-fly \textit{ab-initio}
molecular dynamics \cite{MTM99,DM02a}.

\subsubsection{van-Vleck-based semiclassical approximation}
\label{vvprop}

The Eqs.~(\ref{wigvleckprop}, \ref{wigvleckaction}) defining the
semiclassical Wigner propagator in van Vleck approximation translate
into the following straightforward algorithm to compute the
propagator as such, not operating on any admissible initial Wigner
function:

\renewcommand{\theenumi}{\arabic{enumi}}
\begin{enumerate}
\item \textit{Initial state:} \label{inistate} Define pairs of initial
points ${\bf r}'_{j_\pm}$, $j = 1,\ldots,N$, with common midpoint
${\bf r}' = ({\bf r}'_{j_+} + {\bf r}'_{j_-})/2$, parameterized, e.g.,
by spherical coordinates relative to ${\bf r}'$. A typical value for
the number of classical trajectories, used in most of the calculations
for one-dimensional systems underlying this paper, is $N = 10^6$,
corresponding to $5\times 10^5$ data points available for the final
coarse-graining, step \ref{coarsegrain} below.
\item \textit{Time steps:}
\label{timestep} Realize the
integration over time as a sequence of $L$
steps $t_{l-1}\to t_{l}$, $l= 1,\ldots,L$, $t_l=t'+l\Delta t$,
$\Delta t=(t''-t')/L$. Update the basic ingredients of the
propagator (\ref{wigvleckprop}, \ref{wigvleckaction}) as follows:
\begin{enumerate}
\item \textit{Trajectories} ${\bf r}_{j}(t_{l})$,
$j = 1,\ldots,N,$ according to the classical force field,
\begin{equation}\label{incretra}
\Delta{\bf r}_{j}(t_l)=\mathsf{J}^{\rm t}\nabla
H[{\bf r}_{j}(t_{l})]\Delta t.
\end{equation}
\item \textit{Stability matrices} according to the evolution
equation $\dot{\mathsf{M}} = \mathsf{M}\mathsf{J}^{\rm t}\partial^2
H({\bf r})/\partial{\bf r}^2$ (see ,e.g., \cite{Gro06}),
\begin{equation}\label{incremat}
\Delta \mathsf{M}_{j}(t_l)=\mathsf{M}_{j}(t_l)\mathsf{J}^{\rm t}
\frac{\partial^2 H({\bf r}_{j}(t_{l}))}{\partial
{\bf r}^2_{j}(t_l)}\Delta t.
\end{equation}
It suggests itself to implement (a) and
(b) as a single step, merging Eqs.~(\ref{incretra},
\ref{incremat}) into a single system of linear equations.
\item \textit{Actions} $S_{j}$ as (cf.\ Fig.~\ref{ainama})
\begin{equation}
\Delta S_{j}({\bf r}'',{\bf r}') = [{\bf r}_{j_+}(t_l) -
{\bf r}_{j_-}(t_l)] \wedge [\Delta{\bf r}_{j_-}(t_{l}) +
\Delta{\bf r}_{j_+}(t_{l})]/2 \notag\\
- [H({\bf r}_{j_-}(t_l))-H({\bf r}_{j_+}(t_l))]\Delta t.
\end{equation}
\end{enumerate}
\item \textit{Final state:} \label{finalstate}
\begin{enumerate}
\item Separate elliptic from hyperbolic trajectory pairs according to
\begin{equation}
\hbox{traj.~pair $j$ is}
\begin{cases}
\mathrm{ell.} & \det[\mathsf{M}_{j_+}-\mathsf{M}_{j_-}] > 0,\\
\mathrm{hyp.} & \det[\mathsf{M}_{j_+}-\mathsf{M}_{j_-}] < 0.\\
\end{cases}
\end{equation}
\item \label{coarsegrain}
Within each of the two sheets, coarse-grain the determinantal
prefactor $|\det[\mathsf{M}_+(t) - \mathsf{M}_-(t)]|^{-1/2}$ and the
action $S(t)$ by suitable binning with respect to ${\bf r}''$. In this
step, a possibly inhomogeneous distribution of the initial points (as,
e.g., for polar coordinates) must be accounted for in terms
of weight factors.
\item Calculate the propagator (\ref{wigvleckprop}) for each sheet and
superpose the two contributions.
\end{enumerate}
\end{enumerate}

\subsection{Propagating smooth localized initial states: towards Monte
Carlo algorithms}
\label{montecarlo}

For the more common task of propagating well-localized but
quantum-mechanically admissible initial states (e.g., Gaussians), the
method described in the previous subsection is not optimal.
We can take advantage of the fact that a common midpoint of all
trajectory pairs is not specified, by evaluating all the $N(N-1)/2$
pairs formed by a set of $N$ classical trajectories to gain a factor
$\O(N)$ in efficiency. We have to take into account, however, that the
distribution of centers $\bar{\bf r}_{jk} = ({\bf r}_j+{\bf r}_k)/2$ of
an ensemble of ``satellite'' phase-space points ${\bf r}_j$,
distributed at random or on an ordered grid with probability density
$p_{\rm sat}({\bf r})$, is not this density but its self-convolution,
$p_{\rm ctr}(\bar{\bf r}) = \int \d^{2f} r\, p_{\rm sat}(\bar{\bf r} -
{\bf r}/2) p_{\rm sat}(\bar{\bf r}+{\bf r}/2)$. For Gaussians
(\ref{wiggauss}) this reduces to a contraction by a factor
$\sqrt{2}$.

\begin{figure}[h!]
\begin{center}
\includegraphics[scale=0.3]{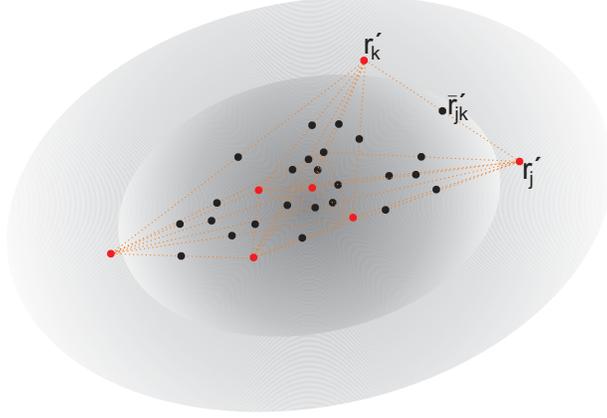}
\caption{Preparation of initial points for the propagation algorithm
for smooth initial distributions, Subsection
\protect\ref{montecarlo}. An ensemble of random ``satellite'' points
${\bf r}'_j$, $j = 1,\ldots,N_{\rm sat}$ (red dots) which serve as
initial points of $N_{\rm sat}$ classical trajectories give rise to
$N_{\rm ctr} = N_{\rm sat}(N_{\rm sat}-1)/2$ midpoint paths $\bar{\bf
r}_{jk}(t)$ starting in the centers $\bar{\bf r}'_{jk} = ({\bf
r}'_j+{\bf r}'_k)/2$ (black dots) and define the support of the
propagator at the final time $t$. The distribution $p_{\rm
ctr}(\bar{\bf r}'_{jk})$ (dark grey) of the centers is that of the
satellites $p_{\rm sat}({\bf r}'_j)$ (light) contracted
by a factor $\sqrt{2}$.
}\label{pucha}
\end{center}
\end{figure}

Accordingly, we propose the following scheme for the propagation of
smooth localized initial states:

\begin{enumerate}
\item \textit{Initial state:} Define a set of initial points
${\bf r}'_{j}$, $j= 1,\ldots,N_{\rm sat}$. A typical value $N_{\rm
sat} = 1000$ now generates $N_{\rm ctr} = 5\times 10^5$ final data
points. This can be done in two ways:
\begin{enumerate}
\item Generate a swarm of random phase-space points
${\bf r}'_{j}$ covering approximately the same phase-space region
as the intended initial Wigner function $W_{\rm ctr}({\bf r}')$,
and associate the weight $W_{\rm ctr}(\bar{\bf r}'_{jk},t')$
to each pair with midpoint $\bar{\bf r}'_{jk}$, $j=
1,\ldots,N_{\rm sat}$, $k= 1,\ldots,j-1$.
\item \label{weightfactors} (For Gaussian initial states only)
Find the distribution $W_{\rm sat}({\bf r}',t')$ with covariance
matrix $\mathsf{A}_{\rm sat} = \mathsf{A}_{\rm ctr}/\sqrt{2}$ that
entails the intended $W_{\rm ctr}(\bar{\bf r}'_{jk},t')$ as its center
distribution and generate the ${\bf r}'_{j}$ according to $W_{\rm
sat}({\bf r}',t')$ (Fig.~\ref{pucha}).
\end{enumerate}
\item \textit{Time steps:} Propagate classically all $N_{\rm sat}$
satellites ${\bf r}'_{j}$ as described in step \ref{timestep} above.
\item \textit{Final state:} Proceed as in step
\ref{finalstate} above for every final midpoint
$\bar{\bf r}''_{jk}$. If option \ref{weightfactors} above has been
chosen, assign the corresponding weights $W_{\rm ctr}(\bar{\bf
r}'_{jk},t')$ to the midpoints $\bar{\bf r}''_{jk}$ in the final
coarse-graining.
\end{enumerate}

Being based on ensembles of random phase-space points distributed
according to some initial density, this propagation scheme
readily integrates in Metropolis-type algorithms. This suggests
itself particularly in the case of high-dimensional spaces where a
direct evaluation of the phase-space integrals involved would be
prohibitive.

\section{Conclusion}
\label{conclusion}
The present work is intended to trigger an interdisciplinary
``technology transfer'', promoting the application in molecular
physics of a semiclassical phase-space propagation scheme that emerged
in quantum chaos. We offer a detailed survey of semiclassical Wigner
propagation, focussing on its implementation and performance as a
numerical tool that provides a natural initial-value
representation. Two complementary approaches to semiclassical
approximations of the Wigner propagator are considered, one based on
the van-Vleck propagator, including nondiagonal contributions to
orbit sums, the other on phase-space path integration
combined with an expansion of the phase. The former possesses a
clear-cut classical skeleton in terms of symplectic geometry, is
surprisingly accurate in significantly nonlinear systems---more so in
the reproduction of the notorious fine nodeline structure of the 
Wigner function than in absolute values---and easily generalizes to 
high-dimensional phase spaces. Two known evils of semiclassical 
propagation, however, return through the back door: caustics, now 
in phase space, and loss of normalization.

The path-integral approach, by contrast, provides a very precise
description at weak yet finite anharmonicity. It resolves
caustics in terms of Airy functions but tends to fail at strong
nonlinearity and does not readily extend to higher dimensions. In
addition, we devise a hybrid based on the
same robust structure of trajectory pairs as the van Vleck approach
but augmented by a uniform approximation to improve the accuracy in
the case of close stationary points. It appears particularly suitable
for heavy-duty numerical applications, as in \textit{ab-initio}
molecular-dynamics simulations \cite{MTM99,DM02a} where integration
times between updates of the potential are relatively short but the
number of freedoms is large and access to second-
and higher-order derivatives of the potential is exceedingly costly.

In view of the relatively little that is known to date about
semiclassical Wigner propagation, our results provide sufficient
qualitative and quantitative evidence to invalidate two popular
connotations: The approach readily captures the time
evolution of quantum coherence effects, including specifically the
propagation of Schr\"odinger-cat states and the reproduction of
tunneling processes. In these cases, it is crucial that even
trajectory pairs with large initial separation be taken into
account. At the same time, we found that the notion of propagating
along any kind of ``quantum trajectory'' is lacking support. While the
deterministic delta function on the classical trajectory that
constitutes the Liouville propagator is replaced by a smooth quantum
spot, no enhancement of this spot on any other deterministic
propagation path is observed.

Various open ends of this work remain to be explored, of which we
mention but a few:

\renewcommand{\theenumi}{\roman{enumi}}
\begin{enumerate}
\item Extending semiclassical Wigner propagation to higher
dimensions. It is significantly facilitated
by the transparent geometric underpinning and by viable
options to adapt it to Monte-Carlo algorithms.

\item Semiclassical Wigner propagation in mixed and even fully chaotic
systems. Recent work in the context of quantum chaos \cite{DP09}
suggests it should work well, but it awaits being tested in
molecular-physics applications.

\item Complexifying phase space. A promising option
how to improve on our trajectory-based semiclassical approximations,
it would resolve caustics and enable a comprehensive description of
tunneling.

\item Including incoherent processes like dephasing and dissipation,
possibly at finite temperature. While this requires major
modifications of semiclassical approximations made for pure states
\cite{GG09}, semiclassical Wigner propagation along trajectory pairs
readily extends to systems with Markovian dissipation
\cite{ORB09}. Alternatively, the Feynman-Vernon influence-functional
theory \cite{FV63,CL83} combines well with phase-space path
integration to achieve high-accuracy semiclassical propagation even in
the presence of memory effects \cite{GSI88}.
\end{enumerate}

\section*{Acknowledgements}
We enjoyed inspiring discussions with Marcus de Aguiar, Frank
Gro\ss mann, Gert Ingold, J\"urgen Korsch, Dominik Marx, Alfredo
Ozorio, Eli Pollak, Uzy Smilansky, and Carlos
Viviescas. Financial aid from Volkswagen Foundation (grant
I/$78\,235$, for TD and EAG), Colciencias (for LAP), Universidad
Nacional de Colombia, and the ALECOL program of the German Academic
Exchange Service DAAD (for EAG) is gratefully acknowledged. We thank
for the hospitality extended to us by the Brazilian Center for Physics
Research CBPF, Rio de Janeiro, MPI for the Physics of Complex Systems,
Dresden, University of Technology Kaisers\-lautern, and University of
Augsburg, where parts of this work were carried out.


\begin{thebibliography}{10}

\bibitem{RG84}
E.~Runge and E.~K.~U. Gross.
\newblock {\em Phys.\ Rev.\ Lett.}, 52:997, 1984.

\bibitem{GBR82}
R.~B. Gerber, V.~Buch, and M.~A. Ratner.
\newblock {\em J.\ Chem.\ Phys.}, 77:3022, 1982.

\bibitem{MMC90}
H.~D. Meyer, U.~Manthe, and L.~S. Cederbaum.
\newblock {\em Chem.\ Phys.\ Lett.}, 165:73, 1990.

\bibitem{SG96}
M.~A. Sep{\'u}lveda and F.~Grossmann.
\newblock {\em Adv.\ Chem.\ Phys.}, 96:191, 1996.

\bibitem{BM72}
M.~V. Berry and K.~E. Mount.
\newblock {\em Rep.\ Prog.\ Phys.}, 35:315, 1972.

\bibitem{Vle28}
J.~H. van Vleck.
\newblock {\em Proc.\ Natl.\ Acad.\ Sci.\ USA}, 14:178, 1928.

\bibitem{Gut67}
M.~C. Gutzwiller.
\newblock {\em J.\ Math.\ Phys.}, 8:1979, 1967.

\bibitem{Lit92}
R.~G. Littlejohn.
\newblock {\em J.\ Stat.\ Phys.}, 68:7, 1992.

\bibitem{Mil74}
W.~H. Miller.
\newblock {\em Adv.\ Chem.\ Phys.}, 25:69, 1974.

\bibitem{LS77}
S.~Levit and U.~Smilansky.
\newblock {\em Ann.\ Phys.\ (N.\ Y.)}, 108:165, 1977.

\bibitem{LM&78}
S.~Levit, K.~M{\"o}hring, U.~Smilansky, and T.~Dreyfus.
\newblock {\em Ann.\ Phys.\ (N.\ Y.)}, 114:223, 1978.

\bibitem{Kla86}
J.~R. Klauder.
\newblock {\em Phys.\ Rev.\ Lett.}, 56:897, 1986.

\bibitem{Mil98}
W.~H. Miller.
\newblock {\em Faraday Discuss.}, 110:1, 1998.

\bibitem{TWM01}
M.~Thoss, H.~Wang, and W.~H. Miller.
\newblock {\em J.\ Chem.\ Phys.}, 114:9220, 2001.

\bibitem{LM06}
J.~Liu and W.~H. Miller.
\newblock {\em J.\ Chem.\ Phys.}, 125:224104, 2006.

\bibitem{Mil01}
W.~H. Miller.
\newblock {\em J.\ Phys.\ Chem.\ A}, 105:2942, 2001.

\bibitem{TW04}
M.~Thoss and H.~Wang.
\newblock {\em Annu.\ Rev.\ Phys.\ Chem.}, 55:299, 2004.

\bibitem{Kay05}
K.~G. Kay.
\newblock {\em Annu.\ Rev.\ Phys.\ Chem.}, 56:255, 2005.

\bibitem{Hel75}
E.~J. Heller.
\newblock {\em J.\ Chem.\ Phys.}, 62:1544, 1975.

\bibitem{HK84}
M.~F. Herman and E.~Kluk.
\newblock {\em Chem.\ Phys.}, 91:27, 1984.

\bibitem{Gro99}
F.~Grossmann.
\newblock {\em Comments At.\ Mol.\ Phys.}, 34:141, 1999.

\bibitem{BA&01}
M.~Baranger, M.~A.~M. de~Aguiar, F.~Keck, H.~J. Korsch, and B.~Schellhaa\ss.
\newblock {\em J.\ Phys.\ A: Math.\ Gen.}, 34:7227, 2001.

\bibitem{Hel91b}
E.~J. Heller.
\newblock {\em J.\ Chem.\ Phys.}, 94:2723, 1991.

\bibitem{Van04}
J.~Van\'{\i}\v{c}ek.
\newblock {\em Phys.\ Rev.\ E}, 70:055201(R), 2004.

\bibitem{Van06}
J.~Van\'{\i}\v{c}ek.
\newblock {\em Phys.\ Rev.\ E}, 73:046204, 2006.

\bibitem{LMV09}
B.~Li, C.~Mollica, and J.~Van\'{\i}\v{c}ek.
\newblock {\em J.\ Chem.\ Phys.}, 131:041101, 2009.

\bibitem{MTM99}
D.~Marx, M.~E. Tuckerman, and G.~J. Martyna.
\newblock {\em Comput.\ Phys.\ Commun.}, 118:166, 1999.

\bibitem{DM02a}
N.~L. Doltsinis and D.~Marx.
\newblock {\em J.\ Theor.\ Comp.\ Chem.}, 1:319, 2002.

\bibitem{CA&09}
M.~Ceotto, S.~Atahan, and A.~Aspuru-Guzik.
\newblock {\em arXiv 0424.0712 [cond.mat.mtrl-sci]}, 2007.

\bibitem{TP09}
J.~Tatchen and E.~Pollak.
\newblock {\em J.\ Chem.\ Phys.}, 130:041103, 2009.

\bibitem{ORB09}
A.~M.~Ozorio de~Almeida, P.~de~M.~Rios, and O.~Brodier.
\newblock {\em J.\ Phys.\ A: Math.\ Theor.}, 42:065306, 2009.

\bibitem{Hus40}
K.~Husimi.
\newblock {\em Proc.\ Phys.\ Math.\ Soc.\ Japan}, 22:264, 1940.

\bibitem{KS85}
J.~R. Klauder and B.~S. Skagerstam.
\newblock {\em Coherent States: Applications in Physics and Mathematical
  Physics}.
\newblock World Scientific, Singapore, 1985.

\bibitem{Kla79}
J.~R. Klauder.
\newblock {\em Phys.\ Rev.\ D}, 19:2349, 1979.

\bibitem{Wig32}
E.~P. Wigner.
\newblock {\em Phys.\ Rev.}, 40:749, 1932.

\bibitem{HO&84}
M.~Hillery, R.~F. O'Connell, M.~O. Scully, and E.~P. Wigner.
\newblock {\em Phys.\ Rep.}, 106:121, 1984.

\bibitem{Sch01}
W.~P. Schleich.
\newblock {\em Quantum Optics in Phase Space}.
\newblock Wiley--VCH, Berlin, 2001.

\bibitem{Ber77}
M.~V. Berry.
\newblock {\em Phil.\ Trans.\ Roy.\ Soc.\ (London) A}, 287:237, 1977.

\bibitem{Zur01}
W.~H. Zurek.
\newblock {\em Nature}, 412:712, 2001.

\bibitem{Hel76}
E.~J. Heller.
\newblock {\em J.\ Chem.\ Phys.}, 65:1289, 1976.

\bibitem{Mcl83}
F.~McLafferty.
\newblock {\em J.\ Chem.\ Phys.}, 78:3253, 1983.

\bibitem{GH95}
M.~Gr{\o}nager and N.~E. Henriksen.
\newblock {\em J.\ Chem.\ Phys.}, 100:5387, 1995.

\bibitem{LS82}
H.~W. Lee and M.~O. Scully.
\newblock {\em J.\ Chem.\ Phys.}, 77:4604, 1982.

\bibitem{Lee92}
H.~W. Lee.
\newblock {\em J.\ Found.\ Phys.}, 22:995, 1992.

\bibitem{Raz96}
M.~Razavy.
\newblock {\em Phys.\ Lett.\ A.}, 119:212, 1996.

\bibitem{DM01}
A.~Donoso and C.~C. Martens.
\newblock {\em Phys.\ Rev.\ Lett.}, 87:223202, 2001.

\bibitem{THW03}
C.~J. Trahan, K.~Hughes, and R.~E. Wyatt.
\newblock {\em J.\ Chem.\ Phys.}, 118:9911, 2003.

\bibitem{MP09}
J.~Moix, E.~Pollak, and J.~Shao.
\newblock {\em Phys.\ Rev.\ A}, 80:052103, 2009.

\bibitem{MS96}
M.~S. Marinov and B.~Segev.
\newblock {\em Phys.\ Rev.\ A}, 54:4752, 1996.

\bibitem{Hel77}
E.~J. Heller.
\newblock {\em J.\ Chem.\ Phys.}, 67:3339, 1977.

\bibitem{Ozo98}
A.~M. \protect{Ozorio de Almeida}.
\newblock {\em Phys.\ Rep.}, 295:265, 1998.

\bibitem{Ber89}
M.~V. Berry.
\newblock {\em Proc.\ Roy.\ Soc.\ Lond.\ A}, 423:219, 1989.

\bibitem{AF98}
O.~Agam and S.~Fishman.
\newblock {\em Phys.\ Rev.\ Lett.}, 73:806, 1998.

\bibitem{TAO01}
F.~Toscano, M.~A.~M. de~Aguiar, and A.~M.~Ozorio de~Almeida.
\newblock {\em Phys.\ Rev.\ Lett.}, 86:59, 2001.

\bibitem{RO02}
P.~P. de~M.~Rios and A.~M.~Ozorio de~Almeida.
\newblock {\em J.\ Phys.\ A: Math.\ Gen.}, 35:2609, 2002.

\bibitem{DVS06}
T.~Dittrich, C.~Viviescas, and L.~Sandoval.
\newblock {\em Phys.\ Rev.\ Lett.}, 96:070403, 2006.

\bibitem{Mar91}
M.~S. Marinov.
\newblock {\em Phys.\ Lett.\ A}, 153:5, 1991.

\bibitem{DP09}
T.~Dittrich and L.~A. Pach\'on.
\newblock {\em Phys.\ Rev.\ Lett.}, 102:150401, 2009.

\bibitem{GT88}
D.~Galetti and A.~F.~R. de~Toledo~Piza.
\newblock {\em Physica}, 149A:267, 1988.

\bibitem{Bal80}
N.~L. Balazs.
\newblock {\em Physica}, 102A:236, 1980.

\bibitem{Fey48}
R.~P. Feynman.
\newblock {\em Rev.\ Mod.\ Phys.}, 20:367, 1948.

\bibitem{Sch81}
L.~S. Schulman.
\newblock {\em Techniques and Applications of Path Integration}.
\newblock Wiley, New York, 1981.

\bibitem{fn1}
If the propagator in original phase space takes the form $G_{\rm
  W}(\bfrho'',\bfrho',t) = g(\bfrho'' - \mathsf{M}(t)\bfrho')$, with
  $g(\bfrho)$ some normalizable distribution and $\mathsf{M}(t)$ a symplectic
  matrix generating a linearized canonical transformation, then the
  Fourier-transformed propagator can be written as $\tilde G_{\rm
  W}(\bfgamma'',t;\bfgamma',0) = (2\pi)^{-1} \tilde
  g(\bfgamma'')\delta(\bfgamma'' - \mathsf{M}(t)\bfgamma')$, where $\tilde
  g(\bfgamma)$ is the Fourier transform of $g(\bfrho)$. This reduces the
  equivalent of Eqs.~(\protect\ref{wigprop},\protect\ref{wigconcat}) in Fourier
  space to multiplications.

\bibitem{AS84}
M.~Abramowitz and I.~A.~Stegun (eds.).
\newblock {\em Pocketbook of Mathematical Functions}.
\newblock Harri Deutsch, Thun, 1984.

\bibitem{VS04}
O.~Vall\'ee and M.~Suares.
\newblock {\em Airy Functions and Applications to Physics}.
\newblock Imperial College Press, London, 2004.

\bibitem{CF&08}
R.~Carles, C.~Fermanian-Kammerer, N.~J. Mauser, and H.~P. Stimming.
\newblock {\em Commun.\ Pure App.\ Anal.}, 8:559, 2009.

\bibitem{KK04}
H.~Kobeissi and M.~Korek.
\newblock {\em Int.\ J.\ Quant.\ Chem.}, 39:23, 2004.

\bibitem{BB06}
O.~Bayrak and I.~Boztosun.
\newblock {\em J.\ Phys.\ A:\ Math.\ Gen.}, 39:6955, 2006.

\bibitem{BL07}
C.~E Burkhardt and J.~J. Leventhal.
\newblock {\em Am.\ J.\ Phys.}, 75:686, 2007.

\bibitem{RG09}
U.~Roy and S.~Ghosh.
\newblock {\em arXiv:0907.3116v1\ [quant-ph]}, 2009.

\bibitem{DS88}
J.~P. Dahl and M.~Springborg.
\newblock {\em J.\ Chem.\ Phys.}, 88:4535, 1988.

\bibitem{HRG04}
C.~Harabati, J.~M. Rost, and F.~Grossmann.
\newblock {\em J.\ Chem.\ Phys.}, 120:26, 2004.

\bibitem{KRO04}
A.~Kenfack, J.~M. Rost, and A.~M.~Ozorio de~Almeida.
\newblock {\em J.\ Phys.\ B: At.\ Mol.\ Opt.\ Phys.}, 37:1645, 2004.

\bibitem{Mor29}
P.~M. Morse.
\newblock {\em Phys.\ Rev.}, 34:57, 1929.

\bibitem{HM&98a}
M.~Hug, C.~Menke, and W.~P. Schleich.
\newblock {\em J.\ Phys.\ A:\ Math.\ Gen.}, 31:L217, 1998.

\bibitem{HM&98bc}
M.~Hug, C.~Menke, and W.~P. Schleich.
\newblock {\em Phys.\ Rev.\ A.}, 57:3188; \textit{ibid.} 3206, 1998.

\bibitem{FR&00}
A.~Frank, A.~L. Rivera, and K.~B. Wolf.
\newblock {\em Phys.\ Rev.\ A.}, 61:054102, 2000.

\bibitem{SK06}
J.~Stanek and J.~Konarski.
\newblock {\em Int.\ J.\ Quant.\ Chem.}, 103:10, 2006.

\bibitem{SB&93}
D.~T. Smithey, M.~Beck, M.~G. Raymer, and A.~Faridani.
\newblock {\em Phys.\ Rev.\ Lett.}, 70:1244, 1993.

\bibitem{Hud74}
R.~L. Hudson.
\newblock {\em Rep.~Math.~Phys.}, 6:249, 1974.

\bibitem{Hel91a}
E.~J. Heller.
\newblock Wavepacket dynamics and quantum chaology.
\newblock In M.-J. Giannoni, A.~Voros, and J.~Zinn-Justin, editors, {\em Chaos
  and Quantum Physics}, volume LII of {\em Les Houches Lectures}, page 547.
  North-Holland, Amsterdam, 1991.

\bibitem{Kay94}
K.~G. Kay.
\newblock {\em J.\ Chem.\ Phys.}, 100:4432, 1994.

\bibitem{VS&01}
L.~M.~K. Vandersypen, M.~Steffen, G.~Breyta, C.~S. Yannon, M.~H. Sherwood, and
  I.~L. Chuang.
\newblock {\em Nature}, 414:883, 2001.

\bibitem{LL65}
L.~D. Landau and I.~M. Lifshitz.
\newblock {\em Quantum Mechanics (Nonrelativistic Theory)}.
\newblock Pergamon, Oxford, 1965.

\bibitem{Cre94}
S.~C. Creagh.
\newblock {\em J.\ Phys.\ A: Math.\ Gen.}, 27:4969, 1994.

\bibitem{XA97}
A.~L.~Xavier Jr. and M.~A.~M. de~Aguiar.
\newblock {\em Phys.\ Rev.\ Lett}, 79:3323, 1997.

\bibitem{BV90}
N.~L. Balazs and A.~Voros.
\newblock {\em Ann.\ Phys. (N.\ Y.)}, 199:123, 1990.

\bibitem{Nov05}
M.~Novaes.
\newblock {\em J.\ Math.\ Phys}, 46:102102, 2005.

\bibitem{MS97b}
M.~S. Marinov and B.~Segev.
\newblock {\em Phys.\ Rev.\ A}, 55:3580, 1997.

\bibitem{JZ&01}
C.~Jung, E.~Ziemniak, M.~Carvajal, A.~Frank, and R.~Lemus.
\newblock {\em Chaos}, 11:464, 2001.

\bibitem{LL92}
A.~J. Lichtenberg and M.~A. Lieberman.
\newblock {\em Regular and Chaotic Dynamics}.
\newblock Number~38 in Applied Mathematical Sciences. Springer, New York, 2nd
  edition, 1992.

\bibitem{AD05}
A.~Arg{\"u}elles and T.~Dittrich.
\newblock {\em Physica A}, 356:72, 2005.

\bibitem{Gro06}
F.~Grossmann.
\newblock {\em J.\ Chem.\ Phys.}, 125:014111, 2006.

\bibitem{GG09}
C.-M. Goletz and F.~Grossmann.
\newblock {\em J.\ Chem.\ Phys.}, 130:244107, 2009.

\bibitem{FV63}
R.~P. Feynman and F.~L. Vernon, Jr.
\newblock {\em Ann.\ Phys.\ (N.\ Y.)}, 24:118, 1963.

\bibitem{CL83}
A.~O. Caldeira and A.~J. Leggett.
\newblock {\em Physica}, 121A:587, 1983.

\bibitem{GSI88}
H.~Grabert, P.~Schramm, and G.-L. Ingold.
\newblock {\em Phys.\ Rep.}, 168:115, 1988.

\end{thebibliography}

\end{document}